\PassOptionsToPackage{unicode}{hyperref}
\PassOptionsToPackage{hyphens}{url}
\PassOptionsToPackage{dvipsnames,svgnames,x11names}{xcolor}
\documentclass[
  10pt,
]{article}
\usepackage{amsmath,amssymb}
\usepackage{lmodern}
\usepackage{iftex}
\ifPDFTeX
  \usepackage[T1]{fontenc}
  \usepackage[utf8]{inputenc}
  \usepackage{textcomp} 
\else 
  \usepackage{unicode-math}
  \defaultfontfeatures{Scale=MatchLowercase}
  \defaultfontfeatures[\rmfamily]{Ligatures=TeX,Scale=1}
\fi
\IfFileExists{upquote.sty}{\usepackage{upquote}}{}
\IfFileExists{microtype.sty}{
  \usepackage[]{microtype}
  \UseMicrotypeSet[protrusion]{basicmath} 
}{}
\makeatletter
\@ifundefined{KOMAClassName}{
  \IfFileExists{parskip.sty}{%
    \usepackage{parskip}
  }{
    \setlength{\parindent}{0pt}
    \setlength{\parskip}{6pt plus 2pt minus 1pt}}
}{
  \KOMAoptions{parskip=half}}
\makeatother
\usepackage{xcolor}
\IfFileExists{xurl.sty}{\usepackage{xurl}}{} 
\IfFileExists{bookmark.sty}{\usepackage{bookmark}}{\usepackage{hyperref}}
\hypersetup{
  pdftitle={Ecosystems are showing symptoms of resilience loss},
  pdfkeywords={cooperation, thresholds, risk, uncertainty, regime
shifts},
  colorlinks=true,
  linkcolor={blue},
  filecolor={Maroon},
  citecolor={blue},
  urlcolor={blue},
  pdfcreator={LaTeX via pandoc}}
\usepackage{graphicx}
\makeatletter
\def\maxwidth{\ifdim\Gin@nat@width>\linewidth\linewidth\else\Gin@nat@width\fi}
\def\maxheight{\ifdim\Gin@nat@height>\textheight\textheight\else\Gin@nat@height\fi}
\makeatother
\setkeys{Gin}{width=\maxwidth,height=\maxheight,keepaspectratio}
\makeatletter
\def\fps@figure{htbp}
\makeatother
\newlength{\cslhangindent}
\newlength{\csllabelwidth}
\newlength{\cslentryspacingunit} 
\newenvironment{CSLReferences}[2] 
 {
  \setlength{\parindent}{0pt}
  \ifodd #1
  \let\oldpar\par
  \def\par{\hangindent=\cslhangindent\oldpar}
  \fi
  \setlength{\parskip}{#2\cslentryspacingunit}
 }%
 {}
\usepackage{calc}

\newcommand{\CSLLeftMargin}[1]{\parbox[t]{\csllabelwidth}{#1}}
\newcommand{\CSLRightInline}[1]{\parbox[t]{\linewidth - \csllabelwidth}{#1}\break}

\usepackage{dcolumn, rotating, longtable, lineno, float, array, tabularx, inputenc}

\graphicspath{{figures/}}
\usepackage[margin=2.5cm]{geometry}
\ifLuaTeX
  \usepackage{selnolig}  
\fi

\title{Ecosystems are showing symptoms of resilience loss}
\author{\small \textbf{Juan C. Rocha}\\
\small Stockholm Resilience Centre\\
\small Stockholm University\\
\small \texttt{\href{mailto:juan.rocha@su.se}{\nolinkurl{juan.rocha@su.se}}}}
\date{}

\begin{document}
\maketitle
\begin{abstract}
Ecosystems around the world are at risk of critical transitions due to
increasing anthropogenic pressures and climate change. Yet it is unclear
where the risks are higher or where in the world ecosystems are more
vulnerable. Here I measure resilience of primary productivity proxies
for marine and terrestrial ecosystems globally. Up to 29\% of global
terrestrial ecosystem, and 24\% marine ones, show symptoms of resilience
loss. These symptoms are shown in all biomes, but Arctic tundra and
boreal forest are the most affected, as well as the Indian Ocean and
Eastern Pacific. Although the results are likely an underestimation,
they enable the identification of risk areas as well as the potential
synchrony of some transitions, helping prioritize areas for management
interventions and conservation.
\end{abstract}

Ecosystems are prone to non-linear dynamics that can shift their
function and structure from one configuration to another (1,2). Examples
of such regime shifts include the transitions from forest to savannas
(3), the collapse of coral reefs (4), kelp forest to urchin barrens (5),
peatland transitions (6), or the emergence of hypoxic dead zones in
coastal systems (7,8). Over 30 different types of regime shifts at the
ecosystem scale have been reported in the literature (9). Yet predicting
when and where they will occur remains challenging for most ecosystems
(10). Understanding this is deeply related to our ability to observe and
measure resilience.

Resilience is the ability of a system to withstand disturbances without
losing its function, structure, and hence its identity (11,12).
Formally, it is the size of the system's basin of attraction (11,13,14).
Several metrics have been used to approximate resilience including the
depth of the basin, slope, distance to the threshold, probability of
tipping, resistance, elasticity, among others (14); yet the simplest and
cross-system indicators used are based on recovery time (11,15,16).
Complex systems when close to critical transitions leave statistical
signatures in the time series of its observables known as \emph{critical
slowing down} (1,17,18). It means that the system takes longer to
recover after a small disturbance, which translates into increases in
variance, autocorrelation, and skewness or flickering (1,10). Similar
indicators exist for spatial data which includes spatial correlations,
discrete Fourier transforms, spatial variance, skewness, power
spectrums, and patch-size distributions (19). These methods however have
some limitations. They require long time series to detect useful signals
(10), and they can fail when regime shifts are driven by stochastic
processes (20,21).

Recent theoretical and empirical developments have addressed some of
these limitations. On the theoretical front, \emph{critical speeding up}
has been proposed as a suitable alternative to detect stochastically
driven critical transitions (22). While critical slowing down relies on
the assumption that resilience loss is driven by a widening and depth
loss of the current basin of attraction, critical speeding up assumes
that the basin shrinks by narrowing the basin (22). Both techniques pick
up resilience loss by measuring changes in the higher moments of the
time series distribution by detecting sudden increase (decrease) of
variance, autocorrelation, skewness or kurtosis. Another proposed proxy
of resilience is the fractal dimension (23,24), which is an indication
of self-similarity across scales. The fractal dimension is related to
how adaptable a system is to perturbations, or how easily it finds modes
to deal with disturbances, applications of this are found in the
diagnosis of cardiac disorders (23,24) and engineering (25). Exit time
has also been proposed as a resilience indicator, it does however
requires high resolution time series with multiple shifts to render
useful insights (16). Autoregressive state-space models and dynamic
linear models are also proposed as alternatives to generic early
warnings which could help circumvent some of their limitations by
calculating changes on the leading eigenvalue in time series, if
\(\lambda >1\) the time series signals a loss of stability and
resilience (26--28). On the empirical realm, recent studies have pointed
to remote sensing products (29) and climate change simulations (30) as
suitable high dimensional datasets for testing some of these tools in
quantifying resilience.

\begin{figure*}[ht]
\centering
\includegraphics[width = 7in, height = 3in]{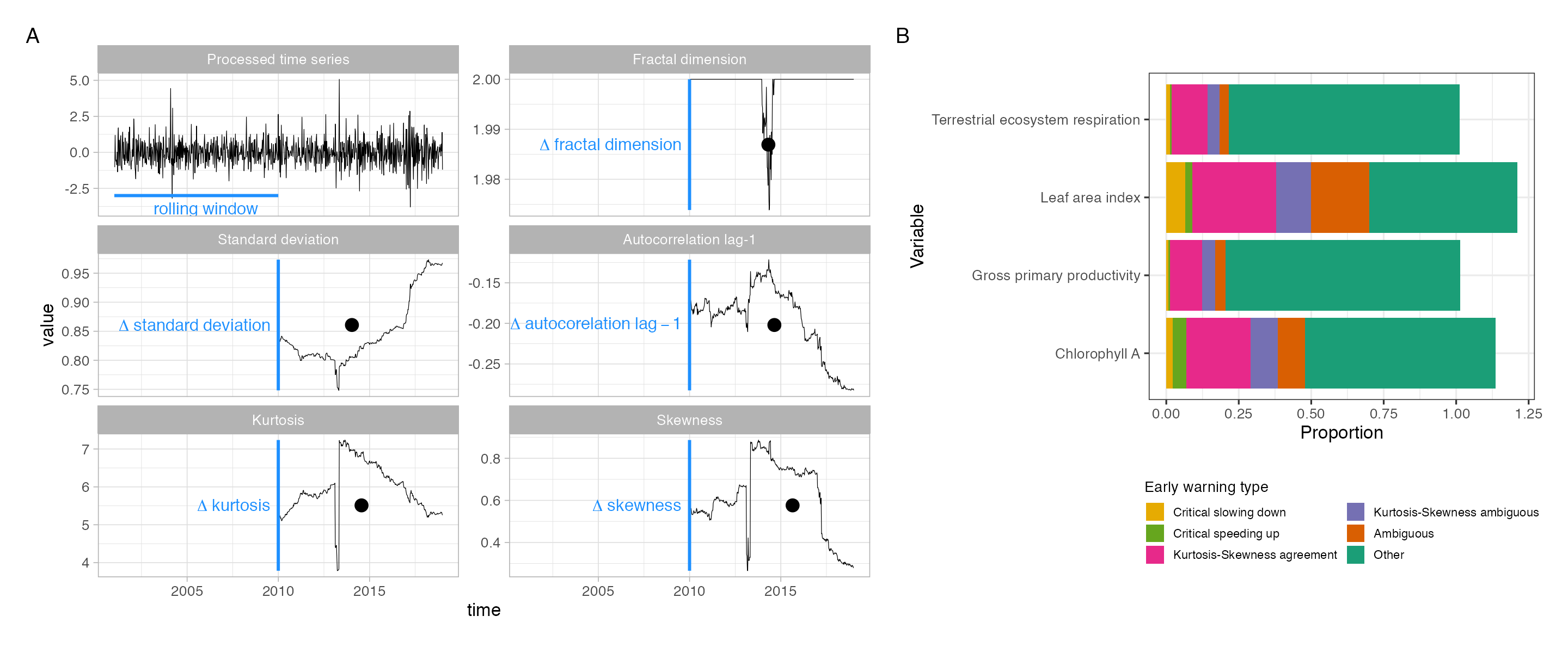}
\caption{\textbf{Example with one pixel} Early warning signals for one pixel of the gross primary productivity dataset. A rolling window of half the length of the time series is used to calculate the dynamic indicators of resilience. $\Delta$ is the difference between maximum and minimum values, and the black points signal the break point of a segmented regression used to detect whether there are big jumps (increase or decrease) on the resilience indicators (A). (B) shows the coherence between resilience indicators across different datasets. They are labelled critical slowing down if both variance and autocorrelations increases, or speeding up if they decrease. If they contradict, they are labelled ambiguous. The same pixels can fill more than one early warning type, thus proportions can be > 1.}
\label{fig:sm-onepxl}
\end{figure*}

This paper aims to identify where regime shifts are likely to occur by
detecting signals of resilience loss in terrestrial and marine
ecosystems primary productivity. It applies the traditional early
warning signals based on critical slowing down; and adapts the methods
to include critical speeding up metrics, and fractal dimension (See
Methods). To that end, \href{http://www.fluxcom.org}{gross primary
productivity}, terrestrial ecosystem respiration, and
\href{https://esa-oceancolour-cci.org}{chlorophyll-a concentration} were
used as proxies of primary productivity of marine and terrestrial
ecosystems. These variables have been harmonized by the
\href{https://www.earthsystemdatalab.net}{Earth System Data Lab}, so all
data layers share the same time (weekly) and spatial (0.25 degree)
resolution (31). To quantify and compare the change in resilience
indicators, the absolute difference between the maximum and minimum
values per indicator was used (\(\Delta\), Fig \ref{fig:sm-onepxl}), and
a segmented regression was used to detect changes in slope and break
points in the time series (Methods). A series of logistic and random
forest regressions were used to gain insights into what is driving
resilience loss in ecosystems world-wide.

\hypertarget{results}{%
\section{Results}\label{results}}

\begin{figure*}[ht]
\centering
\includegraphics[width = 7in, height = 4in]{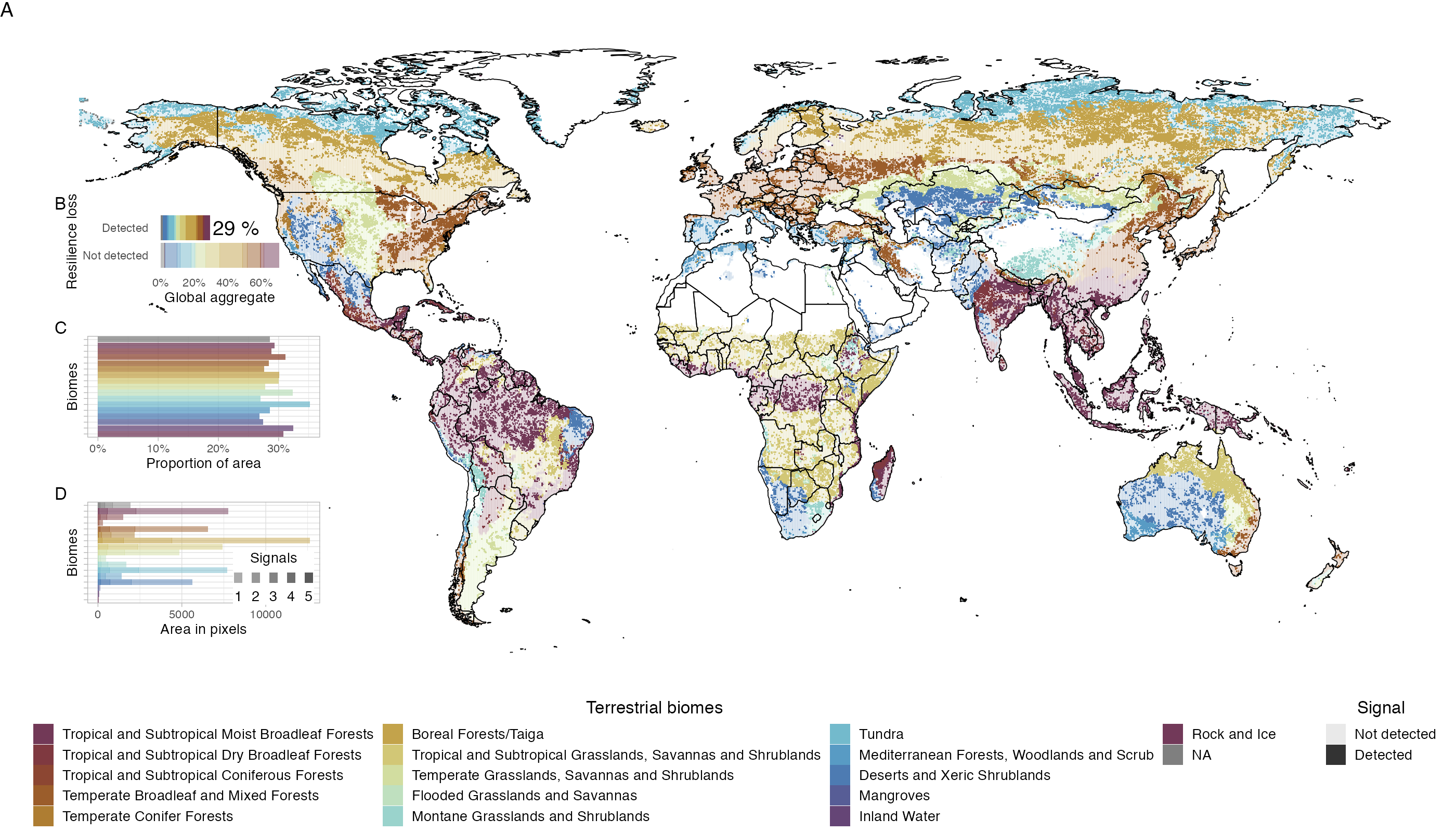}
\caption{\textbf{Resilience loss in terrestrial biomes} Resilience loss for gross primary productivity is approximated as large differences in standard deviation, autocorrelation at lag-1, skewness, kurtosis or fractal dimension. Differences are considered a symptom of resilience loss if they are above the 95\% or below 5\% percentiles of the distribution. A) shows where are biomes showing symptoms of resilience loss, B) shows the global aggregate, C) shows aggregated proportion of area per biome, while D) shows area in 0.25 degree pixels accounting for the number of signals per pixel. A similar figure for terrestrial ecosystem respiration is available in Fig \ref{fig:ter}. Fig \ref{fig:sm-gpp} provides maps for each resilience indicator on the gross primary productivity data, and Fig \ref{fig:sm-ter} on terrestrial ecosystem respiration.}
\label{fig:gpp}
\end{figure*}

The generic resilience indicators do not necessarily align with critical
slowing down or speeding up theories. Fig \ref{fig:sm-onepxl}
illustrates the analysis for one pixel where all indicators show signals
of resilience loss. The time series is first pre-processed to remove
confounding factors such as seasonality or long term oscillations,
log-transformed to reduce the influence of outliers or shock events, and
centered to zero mean and unit variance (Methods). All time series were
first-differenced and passed a root test that guarantees stationarity;
in other words, if a signal is detected it is not the product of
residual long-term trends or seasonal variation on the data. The generic
indicators are then calculated by using a rolling window with length
equivalent to 50\% of the data. To detect changes in trends of the
indicators we used three approaches. First, \(\Delta\) captures the
difference between the minimum and maximum values of the indicator over
time. By itself \(\Delta\) is not very informative unless one has its
distribution for the entire planet to compare against. In the pixel
described in Fig \ref{fig:sm-onepxl} the observed \(\Delta\) is indeed
on the tail of the distribution for all indicators. Second, a linear
regression was fitted to each indicator to test whether the slopes were
different from zero, but it tends to treat sharp increases or decreases
of the resilience indicators as outliers. Hence, the third approach was
a segmented regression, enabling the detection of a point in time where
there is a significant change on the slope of two linear regressions.
The summary of the analysis for all pixels across all datasets shows
that only in few cases autocorrelation and variance increase or decrease
in tandem, the signatures of critical slowing down or speeding up. In
contrast, large values of \(\Delta\), interpreted here as symptoms of
resilience loss, are detected in several different arrangments, most
commonly in the agreement between kurtosis and skewness.

\begin{figure*}[ht]
\centering
\includegraphics[width = 7in, height = 4in]{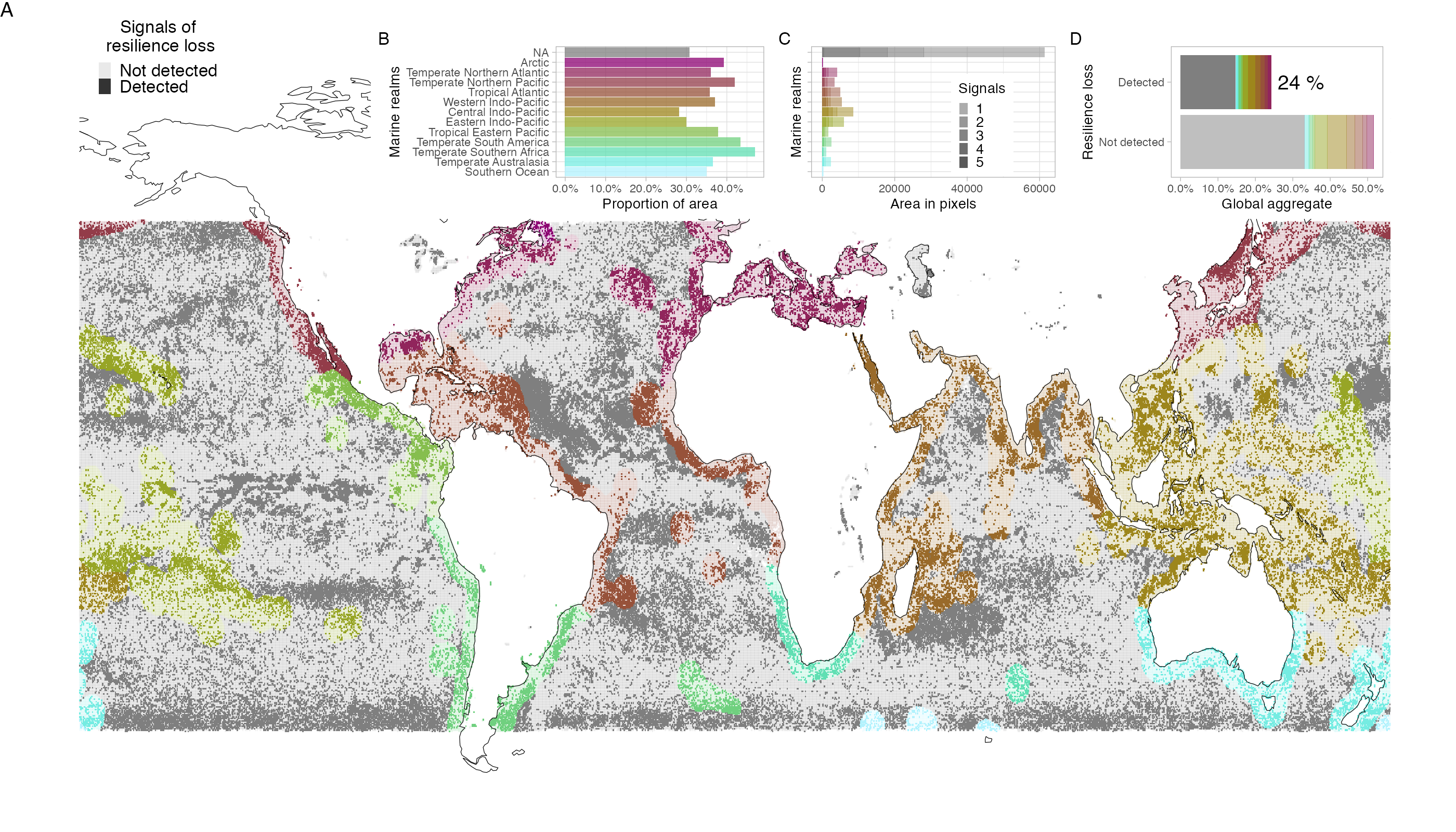}
\caption{\textbf{Resilience loss in marine realms} Detection of resilience loss using chlorophyll-A as proxy of primary productivity (A). B) shows the global aggregate of resilience loss, C) shows aggregated proportion of area per marine realm, while D) shows area in 0.25 degree pixels accounting for the number of signals per pixel. Maps for each resilience indicator are provided in Fig \ref{fig:sm-mar}.}
\label{fig:mar}
\end{figure*}

Ecosystems world-wide are showing symptoms of resilience loss. The
absolute difference in resilience indicators (\(\Delta\)) emphasizes
jumps in the time series and enables comparison with normal variation
adjusted to each biome type. Arctic ecosystems such as boreal forests,
taiga and tundra show the strongest signals of resilience loss globally
(Fig \ref{fig:gpp}, \ref{fig:ter}). However, the extremes of the
distributions (5\% and 95\% percentiles) of each resilience proxy reveal
that all ecosystems are losing resilience, for some of them up to 30\%
of their global area using the gross primary productivity or terrestrial
ecosystem respiration data sets (Fig \ref{fig:sm-gpp},
\ref{fig:sm-ter}). Despite data incompleteness for marine ecosystems at
high latitudes, some signals of resilience loss are detected in Arctic
marine systems and the Southern Ocean (Fig \ref{fig:mar},
\ref{fig:sm-mar}). The Easter Indo-Pacific and Tropical Eastern Pacific
Oceans are the marine realms with larger areas showing symptoms of
resilience loss (Fig \ref{fig:mar}). The high oceans (gray area in Fig
\ref{fig:mar}), however, show by far the larger areas affected with hot
spots outside the Caribbean basin in the Tropical Atlantic, the Tropical
Pacific and southeast Madagascar.

\begin{figure*}[htp]
\centering
\includegraphics[width = 5.5in, height = 3in]{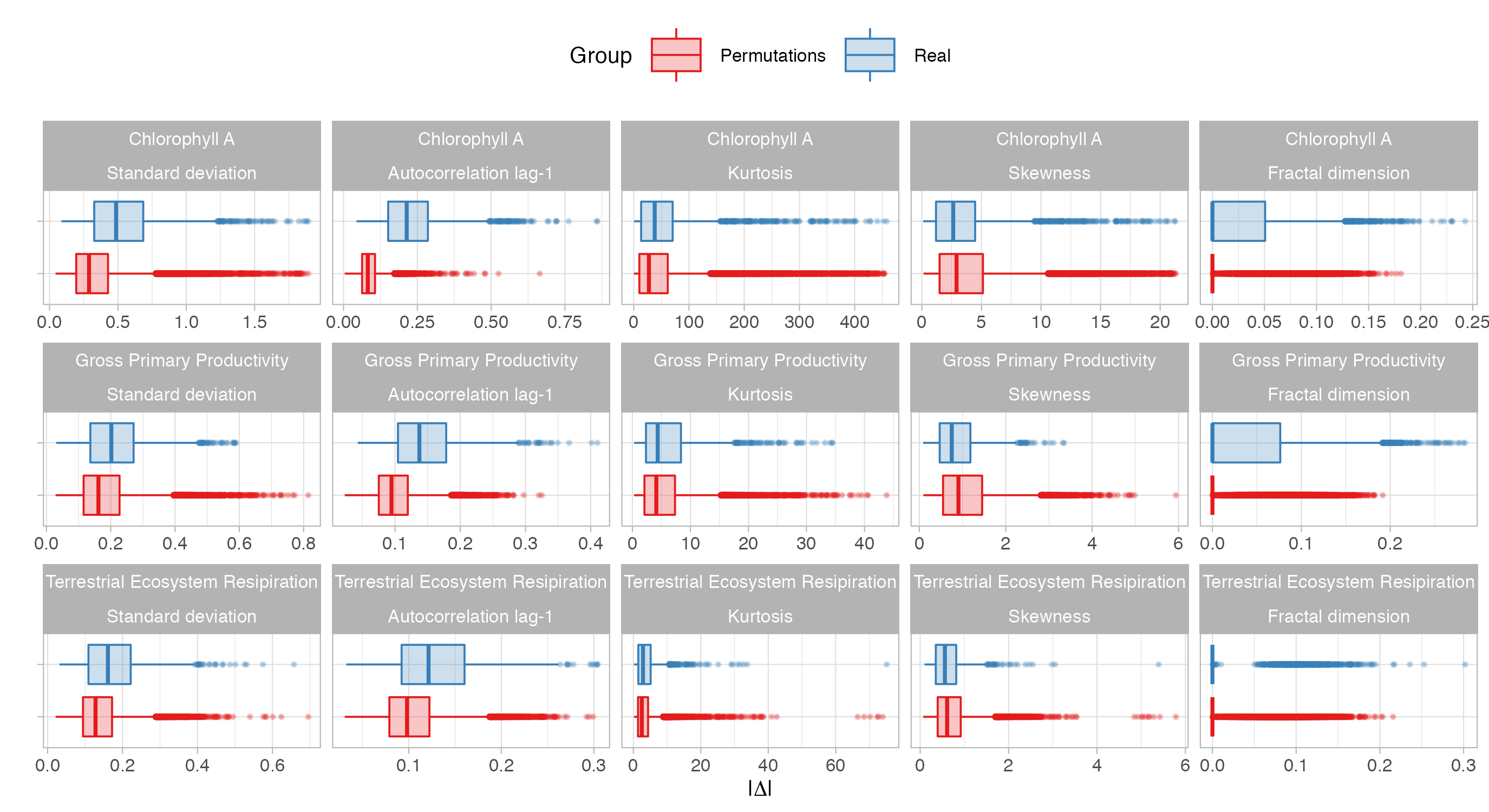}
\caption{\textbf{Permutations tests over time} A sample of 500 pixels where resilience loss was detected by at least three indicators is compared against 5000 permutations of the same time series (10 permutations each). A Wilcoxon two sided test confirms differences in mean between the real and permuted time series (p << 0.05 for all comparisons).}
\label{fig:perms}
\end{figure*}

Symptoms of resilience loss are coherent in space and time. Although the
analysis was done independently for each time series and variable,
spatial aggregation and coincidence of break points in time suggest that
the signals are not artifacts of the data used (Fig \ref{fig:sm-map},
\ref{fig:sm-temp-clus} , \ref{fig:sm-temp-cor}). On the contrary, it
supports the idea that there are edges in the three-dimensional space
(longitude, latitude and time) that enclose volumes whose dynamics can
shift in tandem (30). Resilience indicators are remarkably consistent
across metrics for marine systems both in space (Fig \ref{fig:sm-mar})
and time (Fig \ref{fig:sm-temp-cor}). There is high agreement between
kurtosis and skewness across all data sets (Figs \ref{fig:sm-gpp},
\ref{fig:sm-ter}, \ref{fig:sm-mar}); they can signal the possible
shifting of basin of attraction or dynamic transients (1,10).

\begin{figure*}[ht]
\centering
\includegraphics[width = 7in, height = 2.5in]{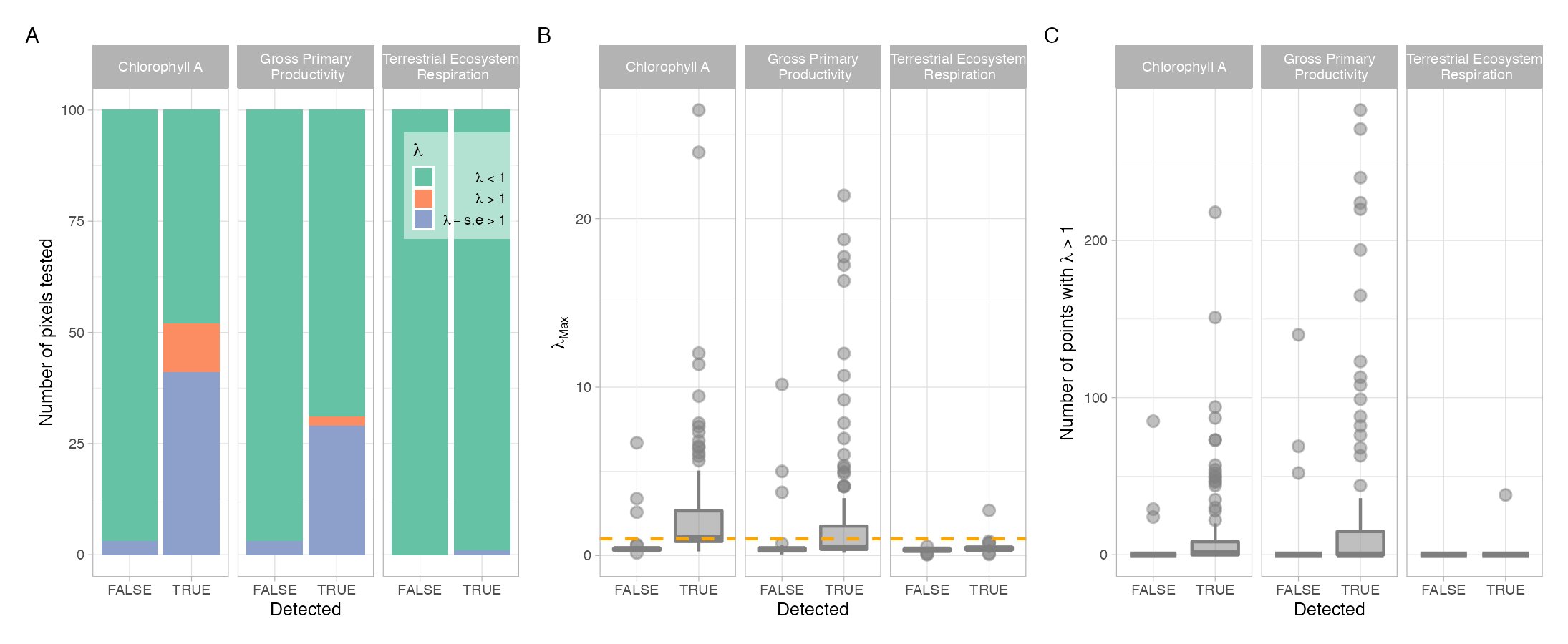}
\caption{\textbf{Model-based indicators} A sample of 100 pixels where resilience loss was detected by at least three indicators is compared against 100 pixels where no signals were detected. The leading eigenvalue of the system is approximated with a dynamic linear autoregressive model}
\label{fig:lambda}
\end{figure*}

An alternative interpretation for coherence of signals is spatial and
temporal autocorrelation, or that they are merely responses to shocks on
environmental variables. Potential issues of temporal autocorrelation
were dealt with by processing the data to the point that the remaining
time series were stationary (32). Issues of spatial autocorrelation are
dealt with (below) by fitting explanatory models and subsampling in
space, while introducing fixed effects for geography related variables
(biome, latitude, longitude). To further test the robustness of the
signals, a subsample of time series of detected places (N = 500) were
compared against the same time series but with the order of observations
changed (10 permutations per time series, 5000 permutations). When the
ordering of observations is lost, the signal of resilience loss
disappears, meaning it has e.g.~a significantly lower autocorrelation
and standard deviation on the absolute value scale (Fig
\ref{fig:perms}). A Wilcoxon two-sided test, comparing the distributions
of permutations agains observed data, suggests that the means are indeed
different (p \textless\textless{} 0.05 for all comparisons). Thus, the
results are not merely noise induced by outliers (e.g.~shocks) because
the same time series with the same outliers but different ordering does
not show the early warnings to the same extent as the observed data. On
the contrary, for some statistics like autocorrelation and standard
deviation, their delta values go back to what is expected to be normal
(the global median).

An additional robustness test is comparing the generic indicators
against model-based early warnings. The latter are autoregressive
dynamic linear models that enable fitting coefficients that change over
time, in particular the approximation of the leading eigenvalue of the
system \(\lambda\) (26--28). For each dataset a sample of a hundred
pixels was randomly drawn for detected places with at least three
different resilience indicators, and compared against places that showed
no early warning (Fig \ref{fig:lambda}). Model-based indicators showed
resilience loss for at least half of the sample of chlorophyll A and a
third of the sample of gross primary productivity. For places where
\(\lambda < 0\) the distribution still shows a displacement towards one
when compared to places with no signals. For terrestrial ecosystem
respiration however, model-based early warnings do not support the
patterns found by the generic indicators.

\begin{figure*}[htp]
\centering
\includegraphics[width = 7in, height = 7in]{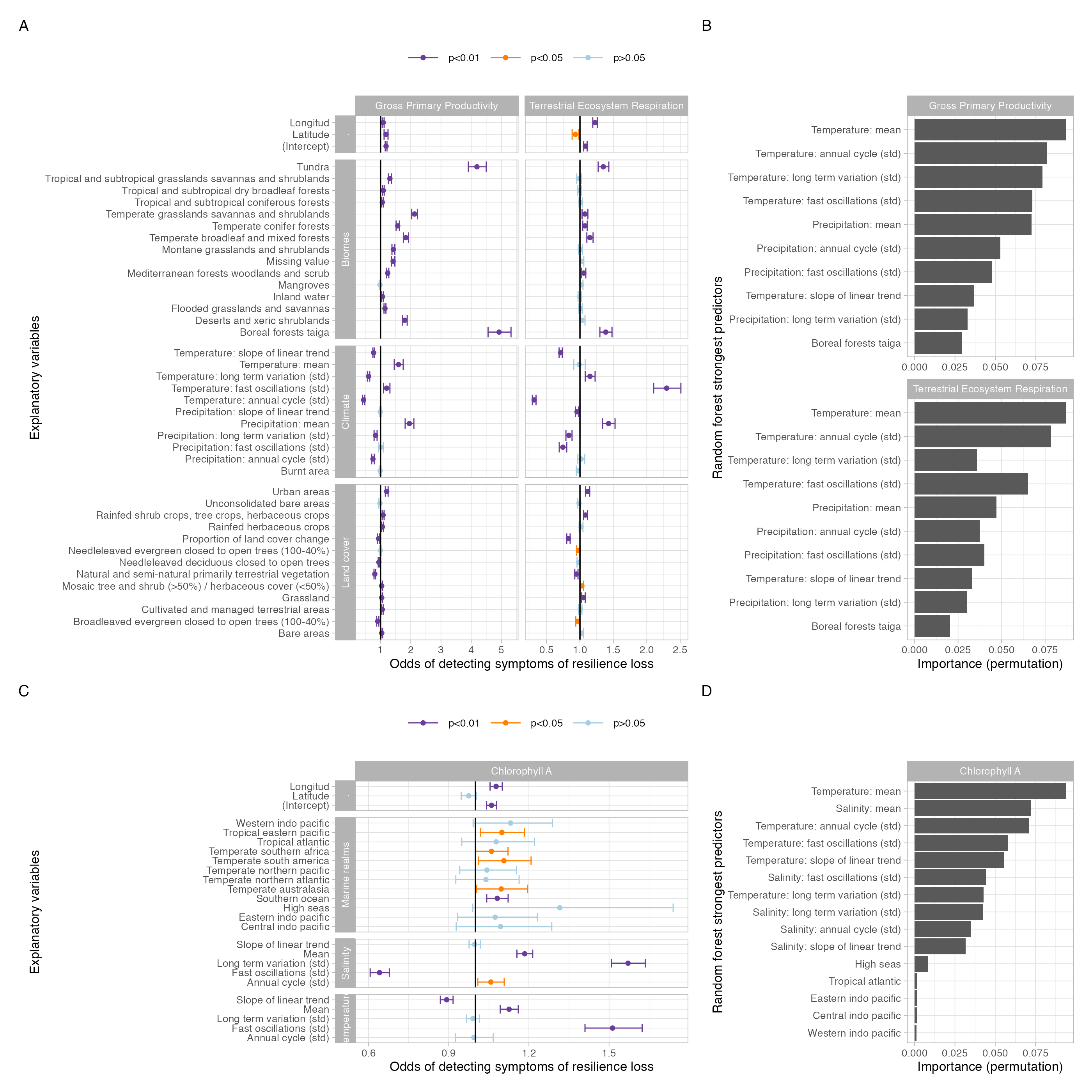}
\caption{\textbf{Predictors of resilience loss} Logistic regressions to predict signals of resilience loss in gross primary productivity, terrestrial ecosystem respiration (A), and chlorophyll A (C). The strongest predictors of the random forest for terrestrial systems (B) and marine realms (D) were calculated with a permutation method. All random forest fitted 1000 trees. The best model for gross primary productivity targeted node size 10 and 12 variables to split at each node (N = 31122, OOB error 0.13), 20 node size and 9 variables for terrestrial ecosystem respiration (N = 29546, OOB error 0.14), and 20 node size and 9 variables for chlorophyll A (N = 54298, OOB error 0.16). }
\label{fig:reg}
\end{figure*}

If resilience is approximated as the size of the basin of attraction,
critical slowing down is sensible to reductions in depth while critical
speeding up is sensible to reductions in width. However, these theories
when applied to ecological problems are typically approached as low
dimensional models with perhaps one controlling factor, one driver.
Recent experimental evidence suggests that when ecosystems are subject
to multiple drivers, early warning indicators can fail or provide
contradictory signals (33). Lack of consistency between signals suggests
that there is no one preferred theory at place. But instead, multiple
drivers are interacting in pushing ecosystems outside their realm of
stability, some of them through increasing stochasticity while others
through slow forcing.

To test that hypothesis, I used logistic regressions and random forests
to investigate what is driving the detection of early warnings.
Detection by at least two proxies of resilience loss was used as a
response variable. Explanatory variables for terrestrial ecosystems
included time series of temperature, precipitation, burned area, and
land cover. For marine systems, the explanatory variables were
restricted to sea surface temperature and sea surface salinity.
Variables with long and frequent time series (temperature, salinity and
precipitation) were filtered with a Fourier transform to test whether
the signals are predicted by long term variation, annual cycles, fast
oscillations, or the linear (slow) trends. All regressions used a
stratified sampling design, balanced on detection, and with fixed
effects per biomes or marine realm respectively (See Method). The
subsampling is necessary to control for two sources of bias. First, it
breaks biases induced by spatial autocorrelation and the potential
inflation of effects or small standard errors (32). Second, it reduces
bias by a relatively larger amount of pixels were resilience loss is not
detected, or by biomes that naturally have larger areas.

The strongest predictors of resilience loss are indeed a combination of
slow forcing and stochasticity in environmental variables such as
temperature, precipitation or sea surface salinity. For terrestrial
biomes, mean temperature, mean precipitation and variability in fast
oscillations and annual cycles are the strongest predictors, while
marine realms are predicted by mean temperature, mean salinity and their
variability at different time scales (Fig \ref{fig:reg}). The logistic
regression facilitates a relatively straightforward interpretation (34).
Once geography and biome type has been controlled for, the reminder
coefficients indicate what increases the odds of detecting resilience
loss. However, this linear approach suffers from correlations in the
predictors, namely the different time scales at which the hypothesis
needs to be tested, whether it is slow forcing or stochasticity at
different time scales what drives the signals. In fact, the predicting
accuracy for the logistic regression is poor (area under the receiver
operating characteristic curve ROC 0.71, 0.65 and 0.59 for gross primary
productivity, terrestrial ecosystem respiration and chlorophyll A
respectively). The random forest approach leads to higher predictive
power (ROC 0.893, 0.874, 0.827 respectively). It is robust to potential
correlations and required less pre-processing for feature engineering
but is less amenable to interpretation (34). It reveals which variables
improve predictions but not necessarily in which direction they are
affecting resilience loss. The results of both approaches confirm the
hypothesis that resilience loss is driven by a combination of slow
forcing and stochasticity on potential drivers.

\hypertarget{discussion}{%
\section{Discussion}\label{discussion}}

Resilience is the ability of any system to deal with change while
keeping its structure and functions, thus its identity (11,12). The
dynamic indicators used here approximate resilience as recovery time
(1,22), and for the fractal dimension a collection of behaviors
available to deal with disturbance (23). These metrics, however, do not
directly inform about other dimensions of resilience such as the size of
the basin of attraction, the amount of disturbance that the system can
stand, the mean return time, the distance to tipping points and
thresholds, or adaptive and transformative capacities (12,14). The
indicators here used can be seen as symptoms of instabilities being
developed on the time series of primary productivity for global
ecosystems over the past two decades. The symptoms do not imply imminent
regime shifts, they help identify places where their probability is
increasing given the limitations of the data. The data is relatively
short to account for critical transitions in the biosphere, yet it is
one of the best observational records to approximate resilience
globally. These symptoms are necessary but not sufficient evidence of
unfolding regime shifts. Yet, the analysis here presented enables the
identification of areas vulnerable to regime shifts at a scale relevant
to decision makers.

Globally 29\% of terrestrial biomes and 24\% of marine realms show
signals of resilience loss in at least one of the indicators used (Fig.
\ref{fig:gpp}, \ref{fig:mar}, \ref{fig:ter}). The overall patterns here
reported agree with recent reports documenting ecosystems' degradation
worldwide. Forests are becoming more vulnerable to droughts, and the
combined effects with increasing fire frequency are exposing them to
major diebacks expected by mid-century (35). Temperature thresholds for
terrestrial primary productivity have been identified (36,37) where
carbon uptake is potentially degraded (sink to source transition). Less
than 10\% of the terrestrial biosphere has already crossed the
threshold, and under business-as-usual scenarios, half of the biosphere
is expected to cross these thresholds by the end of the century, with
the most affected areas being Canadian and Russian boreal taigas as well
as the Amazon and South East Asia rainforests (36). These are biomes
where the strongest signals of resilience loss were detected with a
right shift on the distribution of \(\Delta\). Other studies quantifying
terrestrial ecosystems' resilience with NDVI data also show strong
signals from tundra and boreal forests (38). A recent quantification of
aridity thresholds shows that up to 28.6\% of current dryland area can
cross these thresholds by 2100 in the most drastic climate scenarios
(39). The results here presented confirm early warning signals of
resilience loss in drylands as well, particularly with the fractal
dimension, skewness and kurtosis (Fig. \ref{fig:sm-gpp},
\ref{fig:sm-ter}).

The marine patterns presented here also align with previous studies.
Deepening of the ocean's mixed layer can decrease light conditions near
the surface, decreasing nutrient exchange in the water column and
consequently primary productivity (40). The area outside the Caribbean
basin reported here coincides with an area where salinity has
contributed to ocean stratification (40). Upwelling systems are also
hotspots where resilience loss is identified. Upwelling strength is
expected to change with climate change, with strengthening already
reported in the California and Benguela currents, while weakening in the
Iberian-Canary system (41). Upwelling weakening can limit nutrients in
marine food webs, while strengthening can over enrich nutrients and
facilitate the onset of oxygen minimum zones (7). This study provides an
additional line of support that these systems are being destabilized.

The results are limited by the temporal and spatial scale of the
available data. If the grain of the data is not frequent enough to match
fast dynamics, long enough to capture change in slow processes
(1,10,16), or the spatial resolution is too coarse, local transitions in
space and time can be missed. This is the case with hypoxic areas, where
large oxygen minimum zones are identified, but not the diverse range of
smaller local cases that have been previously reported (7). Only regime
shifts that are drastic enough to change primary productivity as
observed from remote sensing products can be identified. Regime shifts
that impact specific populations or community assemblies without
changing primary productivity are missed. Such is the case of coral
transitions in the Great Barrier Reef where \textasciitilde50\% of the
reef community collapsed following the heatwave events of 2016-17 (42).

Because of these limitations, the results likely are an underestimation.
The estimates are also conservative, with an arbitrary 5\% and 95\%
quantile of the \(\Delta\) distribution as detection threshold. A lower
cutoff would enlarge the areas where resilience loss is detected, but
also increase the risk of false positives. A similar study using NDVI
data estimated up to \textasciitilde65\% of terrestrial ecosystems show
early warning of critical transitions, with strong bias towards boreal
forest and taiga (38). The estimates presented here complement previous
efforts (29,38) in taking into consideration fixed effects by biome and
a pre-processing technique that removes seasonality and long-term
variations that can lead to errors, or bias towards high variable
environments (higher latitudes).

Another limitation of this study is the lack of ground truthing. The
results provide a spatial prediction of where ecosystems might be losing
their resilience. But since critical transitions have not happened yet
in many of these places, it is not possible to contrast the models
presented here with their real predicting power. Current databases that
track such transitions are biased to studies and observations on the
global north and coastal ecosystems (7,9), their coverage is not yet
sufficient to be used as ground truth across all ecosystems and biomes.
Nevertheless, the robustness of the signals detected were tested against
null permutation models and model-based early warnings. The first test
showed the signals are not an artifact of the data, potential temporal
autocorrelation was dealt with stringent data pre-processing and
randomizing the ordering of observations. The second test supported our
findings for chlorophyll A and gross primary productivity, but failed in
supporting the findings of terrestrial ecosystem respiration.

Model-based approaches such as autoregressive time varying models can
help interpreting the results and robustness of generic indicators
(26--28). In this study, these methods gave stronger confidence to the
results from gross primary productivity and less support to terrestrial
ecosystem respiration derived signals, although both results
qualitatively point to the same patterns. The combination of methods
thus helped distinguish what are optimal observables of resilience loss,
an open question and possibly a fruitful avenue for future research
(10). Unfortunately, model-based methods are computationally expensive
and do not scale up to the data requirements of remote sensing products,
on the order of \(10^5\) to \(10^7\) spatial pixels (depending on
spatial resolution) times time series length. Generic indicators are
still useful in narrowing down the scope at which computationally
expensive methods become practical.

Computing the probability density function of \(\Delta\) helps
interpreting signals of resilience loss. Previous work using generic
indicators typically have ground truth in the form of modeling
simulations with some noise or actual experiments. Because the ground
truth is known, there was not concern whether the magnitude of the
increase (decrease) in early warnings was big enough to actually be
considered a warning. Here we do not have ground truth like experiments,
but we can compute the distribution of each indicator for the entire
planet enabling the comparison of what constitutes a big jump on an
indicator as opposed to its normal expected variability for each biome.
The magnitude depends on the indicator and data pre-processing choices,
but the position of the pixel in the distribution is relatively
unaffected. Thus, \(\Delta\) is less sensitive to pre-processing
choices. While in theory increases or decreases on a particular
statistic can be interpreted as resilience loss (an apparent
contradiction), the distribution of \(\Delta\) helps interpreting
whether an increase or decrease is towards the tails of the distribution
(resilience loss) as opposed to an increase or decrease towards the
center of the distribution (a resilience gain). The fractal dimension
here introduced is unambiguous and do not depend on scale, for time
series it is bounded between 1 and 2 and higher values are always more
resilient (23,43). Future studies can use these two facts to track
ecological recovery.

\begin{figure*}[ht]
\centering
\includegraphics[width = 5in, height = 4in]{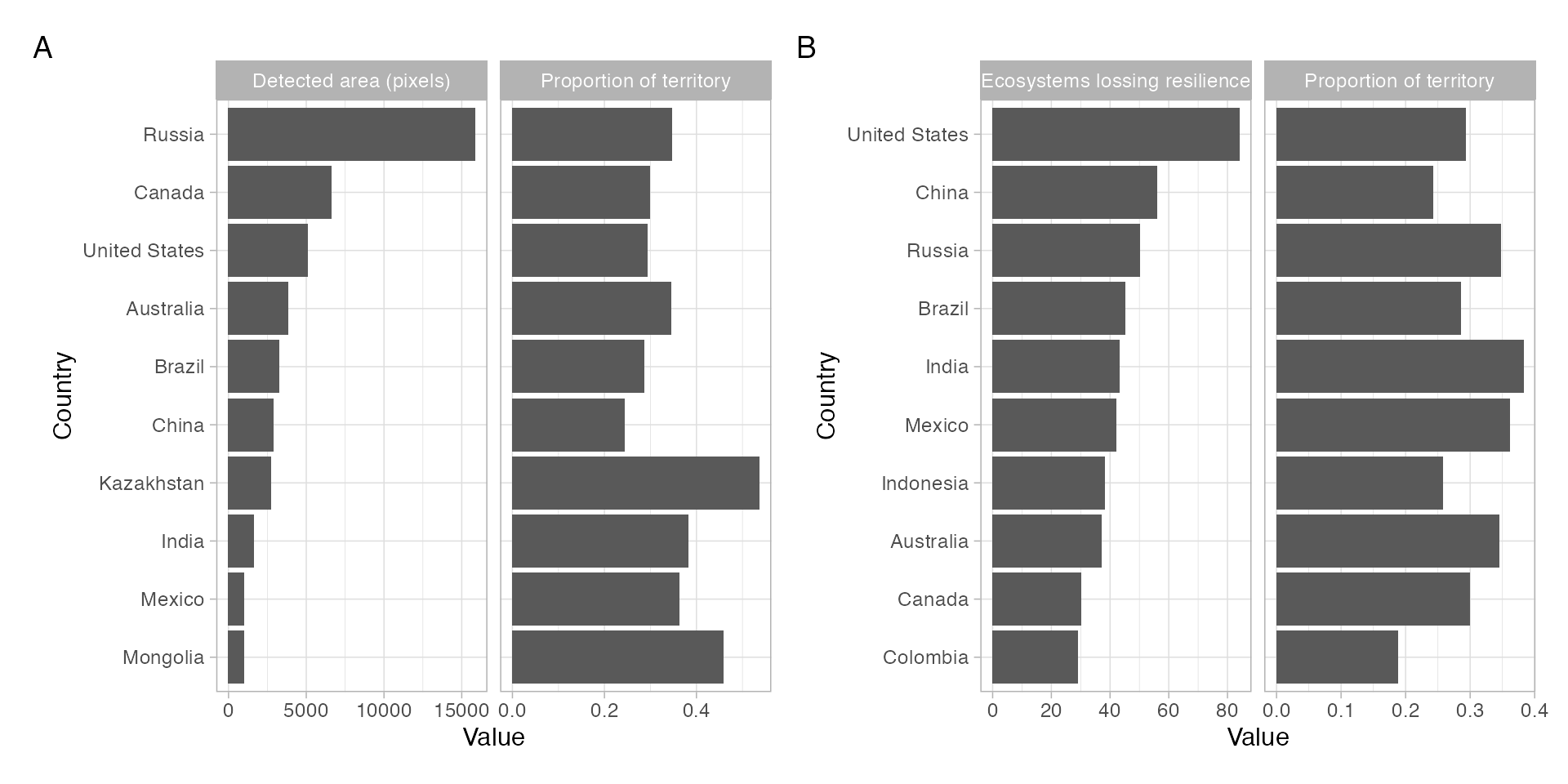}
\caption{\textbf{Most affected countries} Top 10 countries by aggregated area in 0.25 degree pixels showing symptoms of resilience loss and as proportion of their territory (A). Countries ranked in (B) by the number of unique ecosystems showing symptoms of resilience loss and their proportion of territory impacted. }
\label{fig:sm-countries}
\end{figure*}

Despite its limitations, this first-order approximation to resilience
loss can help outline priority areas for management. Russia, Canada, the
US, and Australia are countries with the largest areas of resilience
loss identified, yet by proportion of territory, small island states top
the ranking. When accounting for the diversity of ecosystems showing
signs of resilience loss, megadiverse countries like Brazil, India,
Mexico, Indonesia, Australia, or Colombia are on the top 10 (Fig
\ref{fig:sm-countries}). Australia has recently been reported as a
hotspot for ecological collapses for both marine and terrestrial
ecosystems (44). The spatial resolution of the maps here presented
enable countries, regions, and even municipalities to update land use
planning and take the vulnerability of their ecosystems into
consideration. Companies for example, can take such risk into account
when deciding on relocation, resource outsourcing, or investments.
Countries and municipalities can balance the trade-off between
maximizing a particular ecosystem service (e.g.~a monoculture) in favour
of multifunctional landscapes that include other values such as
recreation, spiritual values or conservation. No matter the scale, we
all have a role to play in caring for ecosystems resilience and
maintaining their ability to provide the ecosystem services we all
depend on.

Future global resilience assessments could benefit from other data
streams, particularly other anthropogenic drivers overlooked here.
Additional data would however require long time series coverage and high
spatial resolution. The accuracy of the predictive models presented here
can be improved by using non-linear approaches such as deep neural
networks or other machine learning techniques (45). However, these
approaches have limited interpretability and often require manually
annotated datasets to measure performance. Qualitative efforts such as
the Global Ocean Oxygen Network (7) or the Regime Shifts Database (9)
can provide the annotated examples to train such models. A spatially
explicit map of resilience loss can also help quantify the risk of
cascading effects in ecosystems previously identified (46). As new Earth
observations become available, these global maps can be updated and
track how ecosystems resilience is evolving, where are they recovering
and where they are becoming more vulnerable. This paper showcases the
first steps toward an ecological resilience observatory.

\hypertarget{methods}{%
\section{Methods}\label{methods}}

\emph{Data:} \href{http://www.fluxcom.org}{Gross primary productivity},
terrestrial ecosystem respiration (both in \(gCm^{-2}d^{-1}\)), and
\href{https://esa-oceancolour-cci.org}{chlorophyll-a concentration}
(\(mgm^{-3}\)) were used as proxies of primary productivity of
terrestrial and marine ecosystems respectively (47--49). Although these
data sets are freely available through the FLUXCOM initiative
(\url{http://www.fluxcom.org}) or the European Space Agency Climate
Change initiative (\url{https://climate.esa.int/}), the versions of the
data used here were harmonized by the
\href{https://www.earthsystemdatalab.net}{Earth System Data Lab} to
weekly observations at 0.25 degree grid resolution, and stored as
\href{https://zarr.readthedocs.io/en/stable/\#}{Zarr} data cubes to
facilitate out of memory computations (31). Gross primary productivity
and terrestrial ecosystem respiration time series span the period
2001-2018 resulting in 210 771 terrestrial pixels with 817 time
observations each, while chlorophyll A span the period 1998-2018
resulting in 418 776 time series with 966 observations each. The Nature
Conservancy classification of terrestrial biomes and ecosystems was used
(16 biomes, 812 ecosystems) based on (50), while marine realms (N=12)
followed (51) classification.

\emph{Pre-processing:} The
\href{https://www.earthsystemdatalab.net}{Earth System Data Lab} is
hosted by the Max Planck Institute of Biogeochemestry and Brockmann
Consult GmbH with the support of the European Space Agency. Their
computing services were used to pre-process the data in their
\texttt{Julia} environment. For each of the resulting time series,
missing data was inputted with the mean seasonal cycle. A fast Fourier
transform was used to filter the time series and remove the trend,
annual cycle, and long-term variability. The remaining fast oscillations
were log-transformed and normalized to zero mean and unit variance. The
cleaned data was exported and further statistical analysis performed in
\texttt{R}. A unit root test showed that a few time series were not
stationary after the pre-processing steps described. Thus all time
series were first-differenced to remove any remaining seasonality.

\emph{Proxies of resilience:} At this point all time series are expected
to have zero mean and unit variance. Resilience loss is here detected by
measuring critical slowing down or speeding up in terms of sharp
increase (decrease) of variance measured as standard deviation,
autocorrelation coefficient at lag-1; or proxies of flickering such as
skewness, and kurtosis. Fractal dimensions were calculated using the
madogram method (43). All these statistics were calculated in rolling
windows half of the size of the time series available. \(\Delta\) is the
difference between maximum and minimum values of the resilience
indicators considering time ordering (thus allowing negative values).
For standard deviation and autocorrelation at lag-1, a positive
\(\Delta\) can indicate critical slowing down or speeding up if
negative. The value of \(\Delta\) however is not informative by itself.
The magnitude depends on the pre-processing choices. Other researchers
prefer a Gaussian filter, or kernel based methods to pre-process time
series. These methods, however, require arbitrary choices that can
optimize for detection of certain biomes while under-estimating for
others. Because the data analyzed spanned about two decades, care should
be taken to avoid bias on the resilience indicator by seasonal
variability, annual cycles or multidecadal oscillations. For these
reasons the Fourier transform was an ideal filter for this study,
returning zero mean and unit variance fast oscillations regardless of
whether the time series comes from a strong seasonal biome (e.g.~boreal
forest) or weak seasonality (e.g.~tropical ones). The value of
\(\Delta\) varies, however, with window size, the smaller the rolling
window, the larger \(\Delta\) becomes. What matters, however, is not the
absolute value of \(\Delta\) but its position relative to the
distribution for the planet, and in particular, relative to the biome
described (due to the seasonality differences). \(\Delta\) has a bimodal
distribution (because zero differences are very unlikely), and outliers
with respect to each biome distribution were reported as places showing
symptoms of resilience loss. Outliers are here defined as places where
\(\Delta\) is unusually extreme, either above the 95\% or below the 5\%
quantiles of \(\Delta\) distribution.

To check for the spatial and temporal coherence across time series, a
segmented regression (52) was used to identify the breaking point at
which the early warning is detected. The segmented regression departed
from a linear fit of the resilience proxy against time, the median of
the indicator was used as starting value, and only one break point was
calculated by default. A Davis tested for the significant difference in
slopes before and after the breaking point. Since the time series are
normalized, the expectation is no difference in variance and hence no
detection of breaking points. If there is a breaking point and the
difference is significant and large, one can expect the signal to be a
warning of resilience loss, specially in the case where other
neighboring areas show similar signals in space and time. The
statistical detection treats each time series independently, but spatial
and temporal coherence of the signal offers supporting evidence for true
detection in the absence of annotated data or ground truth. An
additional line of support can be explored by training models that
exploit information about potential causes of abrupt changes in
ecosystems to predict the detection (or lack) of resilience loss.

\emph{Regressions:} To further test the robustness of the results and
explore what is driving resilience loss, a logistic regression and a
random forest were fitted to classify pixels where at least two metrics
suggested loss of resilience. Explanatory variables included air
temperature at 2m
(\url{https://cds.climate.copernicus.eu/cdsapp\#!/dataset/ecv-for-climate-changee}),
sea surface temperature (53)
(\url{https://climate.esa.int/en/projects/sea-surface-temperature}),
precipitation (\url{https://gpm.nasa.gov/data}), sea surface salinity
(\url{https://climate.esa.int/en/projects/ocean-colour}), burned area
(\url{http://www.globalfiredata.org}), and land cover change
(\url{https://cds.climate.copernicus.eu/cdsapp\#!/dataset/satellite-land-cover}).
For all variables, except land cover, the available data match the
temporal resolution of the data used as proxy of primary productivity,
but not necessarily time span. Thus, a Fourier transform was used to
pre-process the data and separate the linear trend, seasonal cycle,
annual cycle and fast oscillations. For temperature, precipitation and
salinity, their mean value, slope of the linear trend and the standard
deviation of the seasonal cycle, annual cycle and fast oscillations were
used as regressors. This is with the intention of testing whether
resilience loss is driven by variability at different time scales or
changes in slow processes. Land cover data has a higher spatial
resolution (300m) but lower temporal resolution (year). Proportion of
change per land cover class between 1994 and 2018 were calculated and
aggregated at 0.25 degree grid. Burnt area is the aggregated area burned
in hectares during 2001-2018 per pixel.

For logistic regressions, the data was down sampled on the detection
variable (at least two variables indicating resilience loss), after
filtering out rock and ice biomes, log-transforming burnt area,
performing a box-cox transformation on land cover change variables, and
normalizing to zero mean and unit variance all numeric predictors.
Random forest, being more tolerant to variables on their natural units,
were fitted after filtering out rock and ice biomes and down sampling on
detection. 75\% of the data was used for training and tuning hyper
parameters using 10-fold cross validation. All random forests were
fitted with 1000 trees. Best models were assessed against the testing
data (25\%) and variable importance computed with permutation. The best
model for gross primary productivity targeted node size 10 and 12
variables to split at each node (N = 31122, OOB error 0.13), 20 node
size and 9 variables for terrestrial ecosystem respiration (N = 29546,
OOB error 0.14), and 20 node size and 9 variables for chlorophyll A (N =
54298, OOB error 0.16).

\emph{Data availability:} All data used in this study is publicly
available through the Copernicus Climate Change and Atmosphere
Monitoring Service, NASA, FLUXCOM initiative, or the Global fire
emissions database. Links to each data set are provided when introduced
under the section data or regressions.

\emph{Computer code:} All code used in this analysis is available at
\url{https://github.com/juanrocha/ESDL}.

\hypertarget{acknowledgements}{%
\section{Acknowledgements}\label{acknowledgements}}

This work would have not been possible without the open data provided by
Copernicus Climate Change and Atmosphere Monitoring Service, NASA,
FLUXCOM initiative, and the Global fire emissions database. JCR would
like to thank the European Space Agency and the Max Planck Institute of
Biogeochemistry for an early adopter grant to use the Earth System Data
Lab curated data sets and computational facilities. The manuscript has
benefited from comments from Fabian Dablander, Steven Lade, Thorsten
Blenckner, Megan Meacham and Stephen Carpenter. JCR was supported by
Formas grants 942-2015-731, 2020-00198 and 2019-02316, the latter
through the Belmont Forum.

\hypertarget{references}{%
\section*{References}\label{references}}
\addcontentsline{toc}{section}{References}

\hypertarget{refs}{}
\begin{CSLReferences}{0}{0}
\leavevmode\vadjust pre{\hypertarget{ref-Scheffer:2009p4449}{}}%
\CSLLeftMargin{1. }
\CSLRightInline{Scheffer M, Bascompte J, Brock WA, Brovkin V, Carpenter
SR, Dakos V, et al. {Early-warning signals for critical transitions}.
Nature. 2009;461(7260):53--9. }

\leavevmode\vadjust pre{\hypertarget{ref-Anonymous:2004bq}{}}%
\CSLLeftMargin{2. }
\CSLRightInline{Folke C, Carpenter S, Walker B, Scheffer M, Elmqvist T,
Gunderson L, et al. {Regime shifts, resilience, and biodiversity in
ecosystem management}. Annu Rev Ecol Evol S. 2004 Dec;35(1):557--81. }

\leavevmode\vadjust pre{\hypertarget{ref-Hirota:2011p7120}{}}%
\CSLLeftMargin{3. }
\CSLRightInline{Hirota M, Holmgren M, Nes EH van, Scheffer M. {Global
resilience of tropical forest and savanna to critical transitions.}
Science. 2011 Oct;334(6053):232--5. }

\leavevmode\vadjust pre{\hypertarget{ref-Hughes:2017bs}{}}%
\CSLLeftMargin{4. }
\CSLRightInline{Hughes TP, Barnes ML, Bellwood DR, Cinner JE, Cumming
GS, Jackson JBC, et al. {Coral reefs in the Anthropocene}. Nature. 2017
May;546(7656):82--90. }

\leavevmode\vadjust pre{\hypertarget{ref-Ling:2015dra}{}}%
\CSLLeftMargin{5. }
\CSLRightInline{Ling SD, Scheibling RE, Rassweiler A, Johnson CR, Shears
N, Connell SD, et al. {Global regime shift dynamics of catastrophic sea
urchin overgrazing}. Philos Trans R Soc Lond, B, Biol Sci. 2015
Jan;370(1659):20130269--9. }

\leavevmode\vadjust pre{\hypertarget{ref-Turetsky:2015fn}{}}%
\CSLLeftMargin{6. }
\CSLRightInline{Turetsky MR, Benscoter B, Page S, Rein G, Werf GR van
der, Watts A. {Global vulnerability of peatlands to fire and carbon
loss}. Nature Geoscience. 2015 Jan;8(1):11--4. }

\leavevmode\vadjust pre{\hypertarget{ref-Breitburg:2018iz}{}}%
\CSLLeftMargin{7. }
\CSLRightInline{Breitburg D, Levin LA, Oschlies A, Gr\'{e}goire M, Chavez
FP, Conley DJ, et al. {Declining oxygen in the global ocean and coastal
waters.} Science. 2018 Jan;359(6371):eaam7240. }

\leavevmode\vadjust pre{\hypertarget{ref-Diaz:2008p199}{}}%
\CSLLeftMargin{8. }
\CSLRightInline{D\'{i}az RJ, Rosenberg R. {Spreading Dead Zones and
Consequences for Marine Ecosystems}. Science. 2008 Aug;321(5891):926--9.
}

\leavevmode\vadjust pre{\hypertarget{ref-Biggs:2018hx}{}}%
\CSLLeftMargin{9. }
\CSLRightInline{Biggs R, Peterson G, Rocha J. {The Regime Shifts
Database: a framework for analyzing regime shifts in social-ecological
systems}. Ecology and Society. 2018 Jul;23(3):art9. }

\leavevmode\vadjust pre{\hypertarget{ref-Dakos:2015jt}{}}%
\CSLLeftMargin{10. }
\CSLRightInline{Dakos V, Carpenter SR, Nes EH van, Scheffer M.
{Resilience indicators: prospects and limitations for early warnings of
regime shifts}. Philos Trans R Soc Lond, B, Biol Sci. 2015
Jan;370(1659):20130263--3. }

\leavevmode\vadjust pre{\hypertarget{ref-Holling:1973p6861}{}}%
\CSLLeftMargin{11. }
\CSLRightInline{Holling CS. {Resilience and stability of ecological
systems}. Annual review of ecology and systematics. 1973;4:1--23. }

\leavevmode\vadjust pre{\hypertarget{ref-Anonymous:2016fv}{}}%
\CSLLeftMargin{12. }
\CSLRightInline{Folke C. {Resilience (Republished)}. Ecology and
Society. 2016;21(4):art44. }

\leavevmode\vadjust pre{\hypertarget{ref-Clark:1975vj}{}}%
\CSLLeftMargin{13. }
\CSLRightInline{Clark WC. {Notes on Resilience Measures}. 1975;
Available from: \url{http://pure.iiasa.ac.at/id/eprint/338/}}

\leavevmode\vadjust pre{\hypertarget{ref-Krakovska:2021ea}{}}%
\CSLLeftMargin{14. }
\CSLRightInline{Krakovsk\'{a} H, Kühn C, Longo IP. {Resilience of dynamical
systems}. ArXiv {[}Internet{]}. 2021 May; Available from:
\url{http://arxiv.org/abs/2105.10592v1}}

\leavevmode\vadjust pre{\hypertarget{ref-Scheffer:2009wl}{}}%
\CSLLeftMargin{15. }
\CSLRightInline{Scheffer M. {Critical Transitions in Nature and
Society}. Princeton University Press; 2009. }

\leavevmode\vadjust pre{\hypertarget{ref-Arani:2021iv}{}}%
\CSLLeftMargin{16. }
\CSLRightInline{Arani BMS, Carpenter SR, Lahti L, Nes EH van, Scheffer
M. {Exit time as a measure of ecological resilience.} Science. 2021
Jun;372(6547). }

\leavevmode\vadjust pre{\hypertarget{ref-Strogatz:2014wo}{}}%
\CSLLeftMargin{17. }
\CSLRightInline{Strogatz SH. {Nonlinear Dynamics and Chaos}. Hachette
UK; 2014. (With applications to physics, biology, chemistry, and
engineering). }

\leavevmode\vadjust pre{\hypertarget{ref-Scheffer:2012cta}{}}%
\CSLLeftMargin{18. }
\CSLRightInline{Scheffer M, Carpenter SR, Lenton TM, Bascompte J, Brock
W, Dakos V, et al. {Anticipating Critical Transitions}. Science. 2012
Oct;338(6105):344--8. }

\leavevmode\vadjust pre{\hypertarget{ref-Kefi:2014dl}{}}%
\CSLLeftMargin{19. }
\CSLRightInline{K\'{e}fi S, Guttal V, Brock WA, Carpenter SR, Ellison AM,
Livina VN, et al. {Early Warning Signals of Ecological Transitions:
Methods for Spatial Patterns}. PLoS ONE. 2014 Mar;9(3):e92097. }

\leavevmode\vadjust pre{\hypertarget{ref-Hastings:2018gy}{}}%
\CSLLeftMargin{20. }
\CSLRightInline{Hastings A, Abbott KC, Cuddington K, Francis T, Gellner
G, Lai Y-C, et al. {Transient phenomena in ecology}. Science. 2018
Sep;361(6406):eaat6412. }

\leavevmode\vadjust pre{\hypertarget{ref-Hastings:2010p5336}{}}%
\CSLLeftMargin{21. }
\CSLRightInline{Hastings A, Wysham DB. {Regime shifts in ecological
systems can occur with no warning}. Ecol Lett. 2010 Apr;13(4):464--72. }

\leavevmode\vadjust pre{\hypertarget{ref-Titus:2020hb}{}}%
\CSLLeftMargin{22. }
\CSLRightInline{Titus M, Watson J. {Critical speeding up as an early
warning signal of stochastic regime shifts}. Theor Ecol. 2020
Feb;280(1766):20131372. }

\leavevmode\vadjust pre{\hypertarget{ref-West:2017wo}{}}%
\CSLLeftMargin{23. }
\CSLRightInline{West G. {Scale}. Penguin; 2017. (The universal laws of
life, growth, and death in organisms, cities, and in organisms, cities,
economies, and companies). }

\leavevmode\vadjust pre{\hypertarget{ref-West:2010db}{}}%
\CSLLeftMargin{24. }
\CSLRightInline{West BJ. {Fractal physiology and the fractional
calculus: a perspective}. Front Physiol. 2010;1. }

\leavevmode\vadjust pre{\hypertarget{ref-Pavithran:2021jc}{}}%
\CSLLeftMargin{25. }
\CSLRightInline{Pavithran I, Sujith RI. {Effect of rate of change of
parameter on early warning signals for critical transitions}. 2021
Jan;(1):013116. Available from:
\url{https://arxiv.org/abs/2101.11811v1}}

\leavevmode\vadjust pre{\hypertarget{ref-Ives:2012df}{}}%
\CSLLeftMargin{26. }
\CSLRightInline{Ives AR, Dakos V. {Detecting dynamical changes in
nonlinear time series using locally linear state-space models}.
Ecosphere {[}Internet{]}. 2012 Jun;3(6):art58. Available from:
\url{http://www.esajournals.org/doi/abs/10.1890/ES11-00347.1}}

\leavevmode\vadjust pre{\hypertarget{ref-Taranu:2018dg}{}}%
\CSLLeftMargin{27. }
\CSLRightInline{Taranu ZE, Carpenter SR, Frossard V, Jenny J-P, Thomas
Z, Vermaire JC, et al. {Can we detect ecosystem critical transitions and
signals of changing resilience from paleo-ecological records?} Ecosphere
{[}Internet{]}. 2018 Oct;9(10). Available from:
\url{https://esajournals.onlinelibrary.wiley.com/doi/full/10.1002/ecs2.2438}}

\leavevmode\vadjust pre{\hypertarget{ref-Carpenter:2013cr}{}}%
\CSLLeftMargin{28. }
\CSLRightInline{Carpenter SR, Brock WA, Cole JJ, Pace ML. {A new
approach for rapid detection of nearby thresholds in ecosystem time
series}. Oikos {[}Internet{]}. 2013 Oct;123(3):290--7. Available from:
\url{http://doi.wiley.com/10.1111/j.1600-0706.2013.00539.x}}

\leavevmode\vadjust pre{\hypertarget{ref-Verbesselt:2016hn}{}}%
\CSLLeftMargin{29. }
\CSLRightInline{Verbesselt J, Umlauf N, Hirota M, Holmgren M, Nes EH
van, Herold M, et al. {Remotely sensed resilience of tropical forests}.
Nature Climate Change. 2016 Sep;6(11):1028--31. }

\leavevmode\vadjust pre{\hypertarget{ref-Bathiany:2020cl}{}}%
\CSLLeftMargin{30. }
\CSLRightInline{Bathiany S, Hidding J, Scheffer M. {Edge Detection
Reveals Abrupt and Extreme Climate Events}. J Climate. 2020
Jun;33(15):6399--421. }

\leavevmode\vadjust pre{\hypertarget{ref-Mahecha:2020cl}{}}%
\CSLLeftMargin{31. }
\CSLRightInline{Mahecha MD, Gans F, Brandt G, Christiansen R, Cornell
SE, Fomferra N, et al. {Earth system data cubes unravel global
multivariate dynamics}. Earth System Dynamics. 2020;11(1):201--34. }

\leavevmode\vadjust pre{\hypertarget{ref-ives2021statistical}{}}%
\CSLLeftMargin{32. }
\CSLRightInline{Ives AR, Zhu L, Wang F, Zhu J, Morrow CJ, Radeloff VC.
Statistical inference for trends in spatiotemporal data. Remote Sensing
of Environment. 2021;266:112678. }

\leavevmode\vadjust pre{\hypertarget{ref-Dai:2015dwa}{}}%
\CSLLeftMargin{33. }
\CSLRightInline{Dai L, Korolev KS, Gore J. {Relation between stability
and resilience determines the performance of early warning signals under
different environmental drivers.} P Natl Acad Sci Usa. 2015
Aug;112(32):10056--61. }

\leavevmode\vadjust pre{\hypertarget{ref-Kuhn:2013vn}{}}%
\CSLLeftMargin{34. }
\CSLRightInline{Kuhn M, Johnson K. {Applied Predictive Modeling}.
Springer Science {\&} Business Media; 2013. }

\leavevmode\vadjust pre{\hypertarget{ref-Williams:2013iy}{}}%
\CSLLeftMargin{35. }
\CSLRightInline{Williams AP, Allen CD, Macalady AK, Griffin D.
{Temperature as a potent driver of regional forest drought stress and
tree mortality}. Commun Biol. 2013;3. }

\leavevmode\vadjust pre{\hypertarget{ref-Duffy:2021br}{}}%
\CSLLeftMargin{36. }
\CSLRightInline{Duffy KA, Schwalm CR, Arcus VL, Koch GW, Liang LL,
Schipper LA. {How close are we to the temperature tipping point of the
terrestrial biosphere?} Science Advances. 2021 Jan;7(3):eaay1052. }

\leavevmode\vadjust pre{\hypertarget{ref-Johnston:2021fg}{}}%
\CSLLeftMargin{37. }
\CSLRightInline{Johnston ASA, Meade A, Ardö J, Arriga N, Black A,
Blanken PD, et al. {Temperature thresholds of ecosystem respiration at a
global scale.} Nature Ecology {\&} Evolution. 2021 Apr;5(4):487--94. }

\leavevmode\vadjust pre{\hypertarget{ref-Feng:2021gr}{}}%
\CSLLeftMargin{38. }
\CSLRightInline{Feng Y, Su H, Tang Z, Wang S, Zhao X, Zhang H, et al.
{Reduced resilience of terrestrial ecosystems locally is not reflected
on a global scale}. Commun Biol. 2021;2(88). }

\leavevmode\vadjust pre{\hypertarget{ref-Berdugo:2020kk}{}}%
\CSLLeftMargin{39. }
\CSLRightInline{Berdugo M, Delgado-Baquerizo M, Soliveres S,
Hern\'{a}ndez-Clemente R, Zhao Y, Gait\'{a}n JJ, et al. {Global ecosystem
thresholds driven by aridity}. Science. 2020 Feb;367(6479):787--90. }

\leavevmode\vadjust pre{\hypertarget{ref-Sallee:2021ju}{}}%
\CSLLeftMargin{40. }
\CSLRightInline{Sall\'{e}e J-B, Pellichero V, Akhoudas C, Pauthenet E,
Vignes L, Schmidtko S, et al. {Summertime increases in upper-ocean
stratification and mixed-layer depth.} Nature. 2021
Mar;591(7851):592--8. }

\leavevmode\vadjust pre{\hypertarget{ref-Sydeman:2014dh}{}}%
\CSLLeftMargin{41. }
\CSLRightInline{Sydeman WJ, Garc\'{i}a-Reyes M, Schoeman DS, Rykaczewski RR,
Thompson SA, Black BA, et al. {Climate change and wind intensification
in coastal upwelling ecosystems.} Science. 2014 Jul;345(6192):77--80. }

\leavevmode\vadjust pre{\hypertarget{ref-Hughes:2018ba}{}}%
\CSLLeftMargin{42. }
\CSLRightInline{Hughes TP, Kerry JT, Baird AH, Connolly SR, Dietzel A,
Eakin CM, et al. {Global warming transforms coral reef assemblages}.
Nature. 2018 Apr;556(7702):492--6. }

\leavevmode\vadjust pre{\hypertarget{ref-Gneiting:2012be}{}}%
\CSLLeftMargin{43. }
\CSLRightInline{Gneiting T, Ševč\'{i}kov\'{a} H, Percival D. {Estimators of
fractal dimension: Assessing the roughness of time series and spatial
data}. 2012;27(2):247--77. }

\leavevmode\vadjust pre{\hypertarget{ref-Bergstrom:2021ey}{}}%
\CSLLeftMargin{44. }
\CSLRightInline{Bergstrom DM, Wienecke BC, Hoff J van den, Hughes L,
Lindenmayer DB, Ainsworth TD, et al. {Combating ecosystem collapse from
the tropics to the Antarctic.} Glob Change Biol. 2021
May;27(9):1692--703. }

\leavevmode\vadjust pre{\hypertarget{ref-Reichstein:2019ii}{}}%
\CSLLeftMargin{45. }
\CSLRightInline{Reichstein M, Camps-Valls G, Stevens B, Jung M, Denzler
J, Carvalhais N, et al. {Deep learning and process understanding for
data-driven Earth system science}. Nature. 2019 Feb;566(7743):195--204.
}

\leavevmode\vadjust pre{\hypertarget{ref-Rocha:2018gn}{}}%
\CSLLeftMargin{46. }
\CSLRightInline{Rocha JC, Peterson G, Bodin O, Levin S. {Cascading
regime shifts within and across scales}. Science. 2018
Dec;362(6421):1379--83. }

\leavevmode\vadjust pre{\hypertarget{ref-Jung:2019hr}{}}%
\CSLLeftMargin{47. }
\CSLRightInline{Jung M, Koirala S, Weber U, Ichii K, Gans F, Camps-Valls
G, et al. {The FLUXCOM ensemble of global land-atmosphere energy
fluxes.} Scientific Data. 2019 May;6(1):74. }

\leavevmode\vadjust pre{\hypertarget{ref-bg-13-4291-2016}{}}%
\CSLLeftMargin{48. }
\CSLRightInline{Tramontana G, Jung M, Schwalm CR, Ichii K, Camps-Valls
G, R 'aduly B, et al. {Predicting carbon dioxide and energy fluxes
across global FLUXNET sites with regression algorithms}. BIOGEOSCIENCES.
2016;13(14):4291--313. }

\leavevmode\vadjust pre{\hypertarget{ref-Sathyendranath:2019hs}{}}%
\CSLLeftMargin{49. }
\CSLRightInline{Sathyendranath S, Brewin RJW, Brockmann C, Brotas V,
Calton B, Chuprin A, et al. {An Ocean-Colour Time Series for Use in
Climate Studies: The Experience of the Ocean-Colour Climate Change
Initiative (OC-CCI).} Sensors (Basel). 2019 Oct;19(19). }

\leavevmode\vadjust pre{\hypertarget{ref-Olson:2001ju}{}}%
\CSLLeftMargin{50. }
\CSLRightInline{Olson DM, Dinerstein E, Wikramanayake ED, Burgess ND,
Powell GVN, Underwood EC, et al. {Terrestrial Ecoregions of the World: A
New Map of Life on Earth}. BioScience. 2001;51(11):933. }

\leavevmode\vadjust pre{\hypertarget{ref-Spalding:2007ga}{}}%
\CSLLeftMargin{51. }
\CSLRightInline{Spalding MD, Fox HE, Allen GR, Davidson N, Ferdaña ZA,
Finlayson M, et al. {Marine Ecoregions of the World: A
Bioregionalization of Coastal and Shelf Areas}. BioScience. 2007
Jul;57(7):573--83. }

\leavevmode\vadjust pre{\hypertarget{ref-Muggeo:2003go}{}}%
\CSLLeftMargin{52. }
\CSLRightInline{Muggeo VMR. {Estimating regression models with unknown
break-points}. Stat Med. 2003;22(19):3055--71. }

\leavevmode\vadjust pre{\hypertarget{ref-Merchant:2019fs}{}}%
\CSLLeftMargin{53. }
\CSLRightInline{Merchant CJ, Embury O, Bulgin CE, Block T, Corlett GK,
Fiedler E, et al. {Satellite-based time-series of sea-surface
temperature since 1981 for climate applications.} Scientific Data. 2019
Oct;6(1):223. }

\end{CSLReferences}

\pagebreak

\hypertarget{sec:SM}{%
\section{Supplementary Material}\label{sec:SM}}

\renewcommand\thefigure{S\arabic{figure}}
\renewcommand\thetable{S\arabic{table}}
\setcounter{table}{0}
\setcounter{figure}{0}

\begin{figure*}[ht]
\centering
\includegraphics[width = 7in, height = 4in]{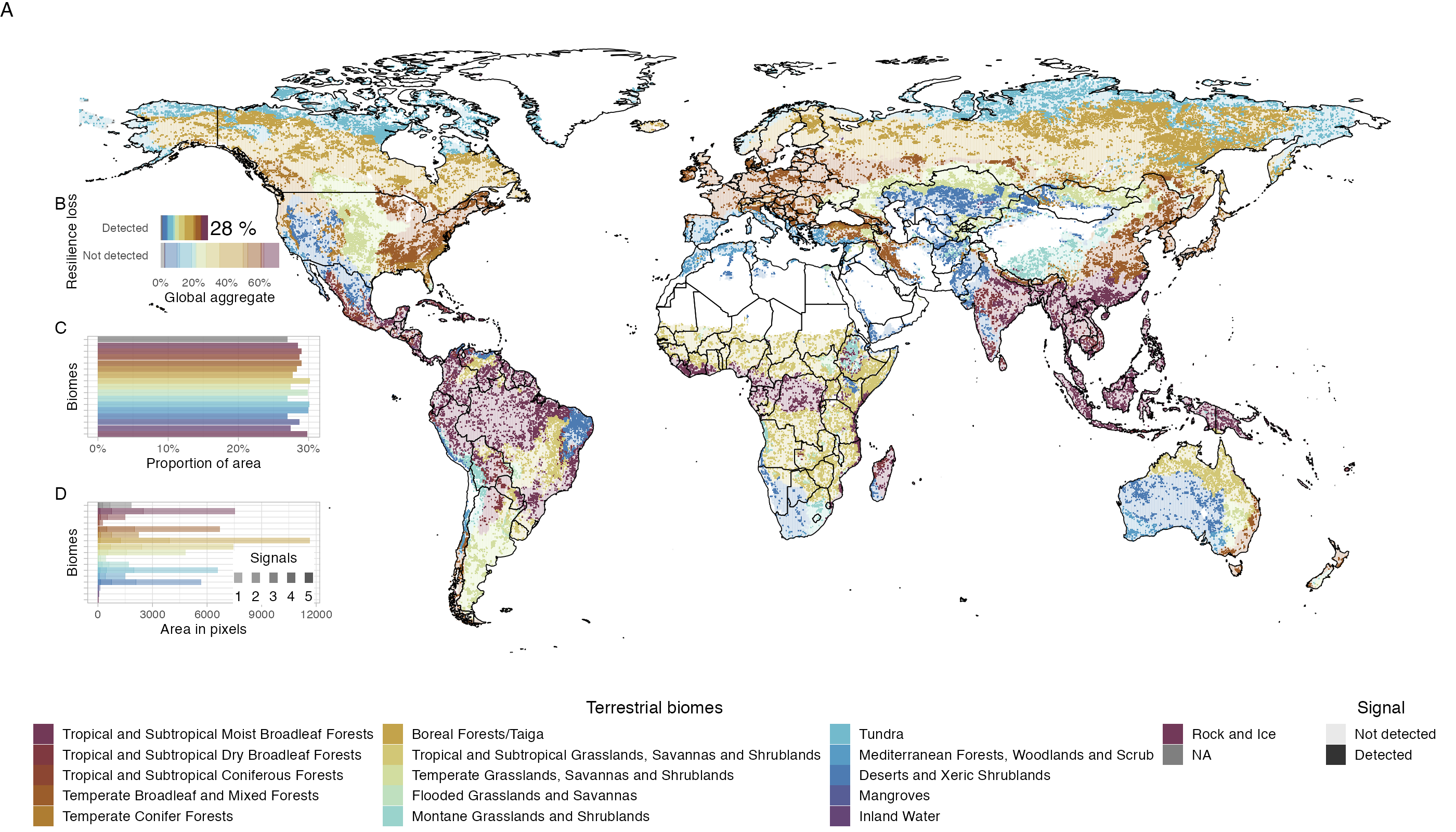}
\caption{\textbf{Terrestrial ecosystem respiration} Detection of resilience loss in terrestrial biomes for terrestrial ecosystem respiration. A) shows where biomes are showing symptoms of resilience loss,  B) shows the global aggregate, C) shows aggregated proportion of area per biome, while C) shows area in 0.25 degree pixels.}
\label{fig:ter}
\end{figure*}

\begin{figure*}[ht]
\centering
\includegraphics[width = 7in, height = 8in]{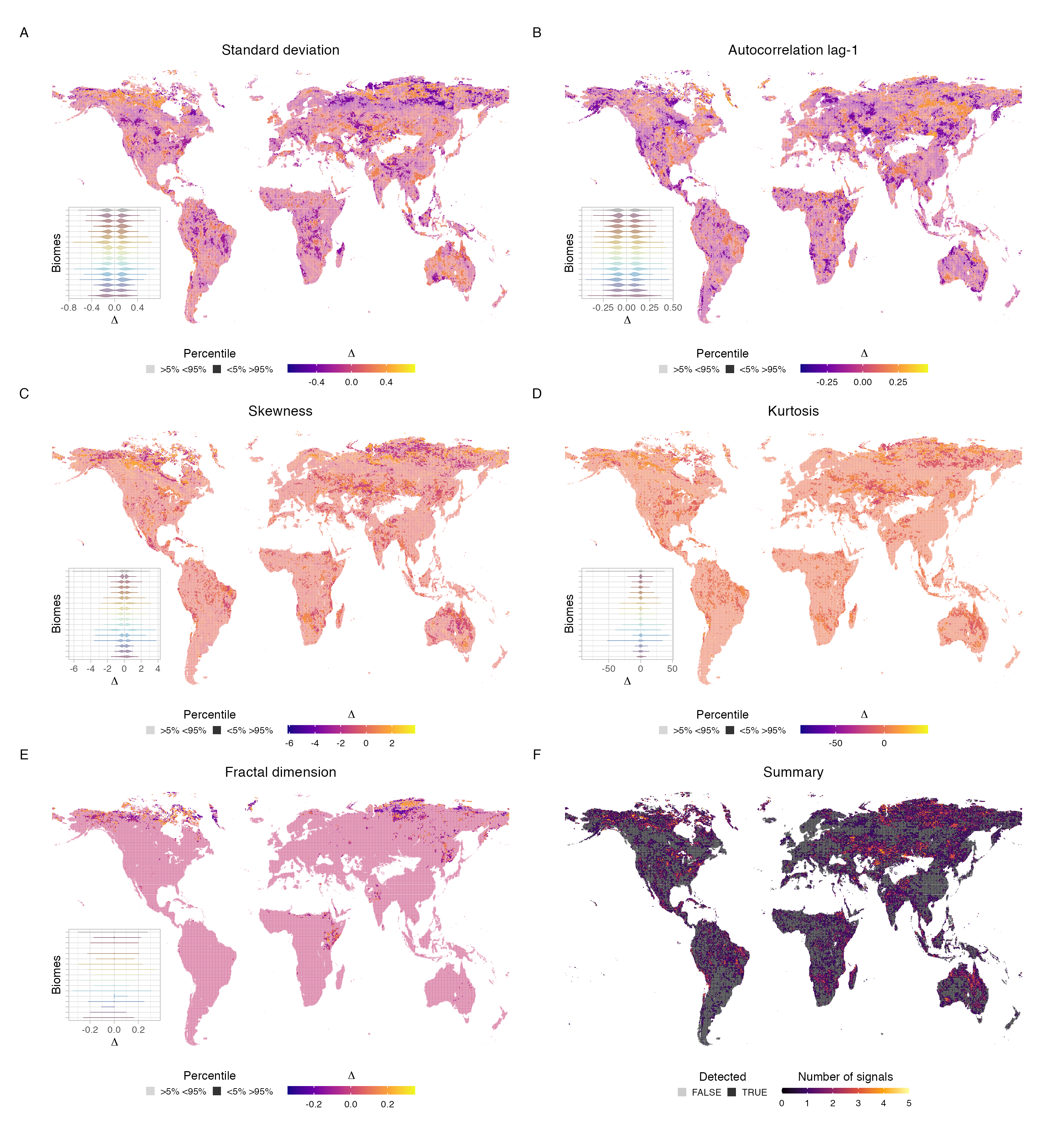}
\caption{\textbf{Resilience indicators in gross primary productivity} Maps of the $\Delta$ outliers per biome are shown for standard deviation (A), autocorrelation at lag-1 (B), skewness (C), kurtosis (D), and fractal dimension (E). Insets show the distribution of $\Delta$ for each biome following the same colouring scheme of fig \ref{fig:gpp}. (F) is the summary with the aggregated number of signals.}
\label{fig:sm-gpp}
\end{figure*}

\begin{figure*}[ht]
\centering
\includegraphics[width = 7in, height = 8in]{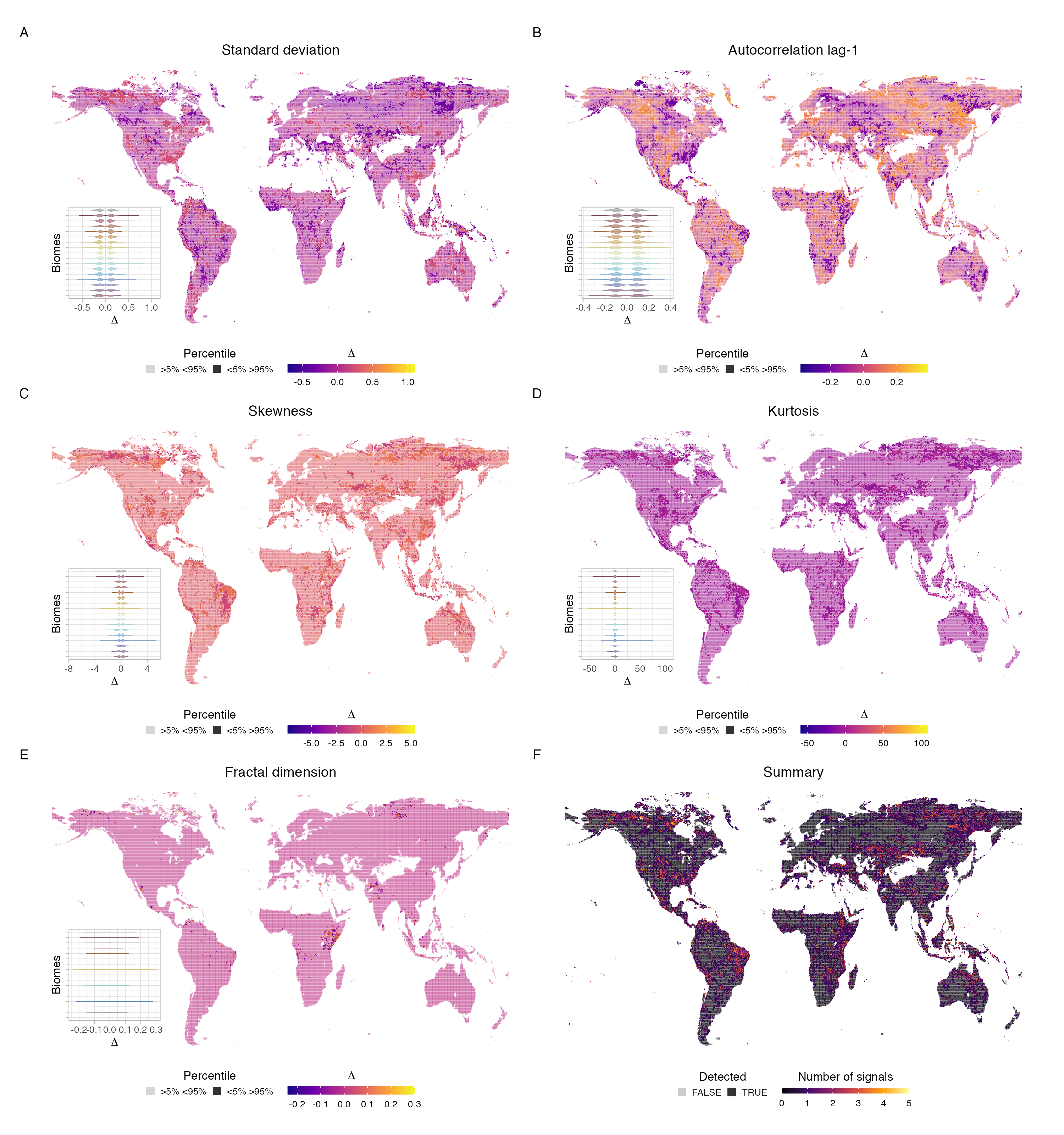}
\caption{\textbf{Resilience indicators in terrestrial ecosystem respiration} Maps of the $\Delta$ outliers per biome are shown for standard deviation (A), autocorrelation at lag-1 (B), skewness (C), kurtosis (D), and fractal dimension (E). Insets show the distribution of $\Delta$ for each biome following the same colouring scheme of fig \ref{fig:ter}. (F) is the summary with the aggregated number of signals.}
\label{fig:sm-ter}
\end{figure*}

\begin{figure*}[ht]
\centering
\includegraphics[width = 7in, height = 8in]{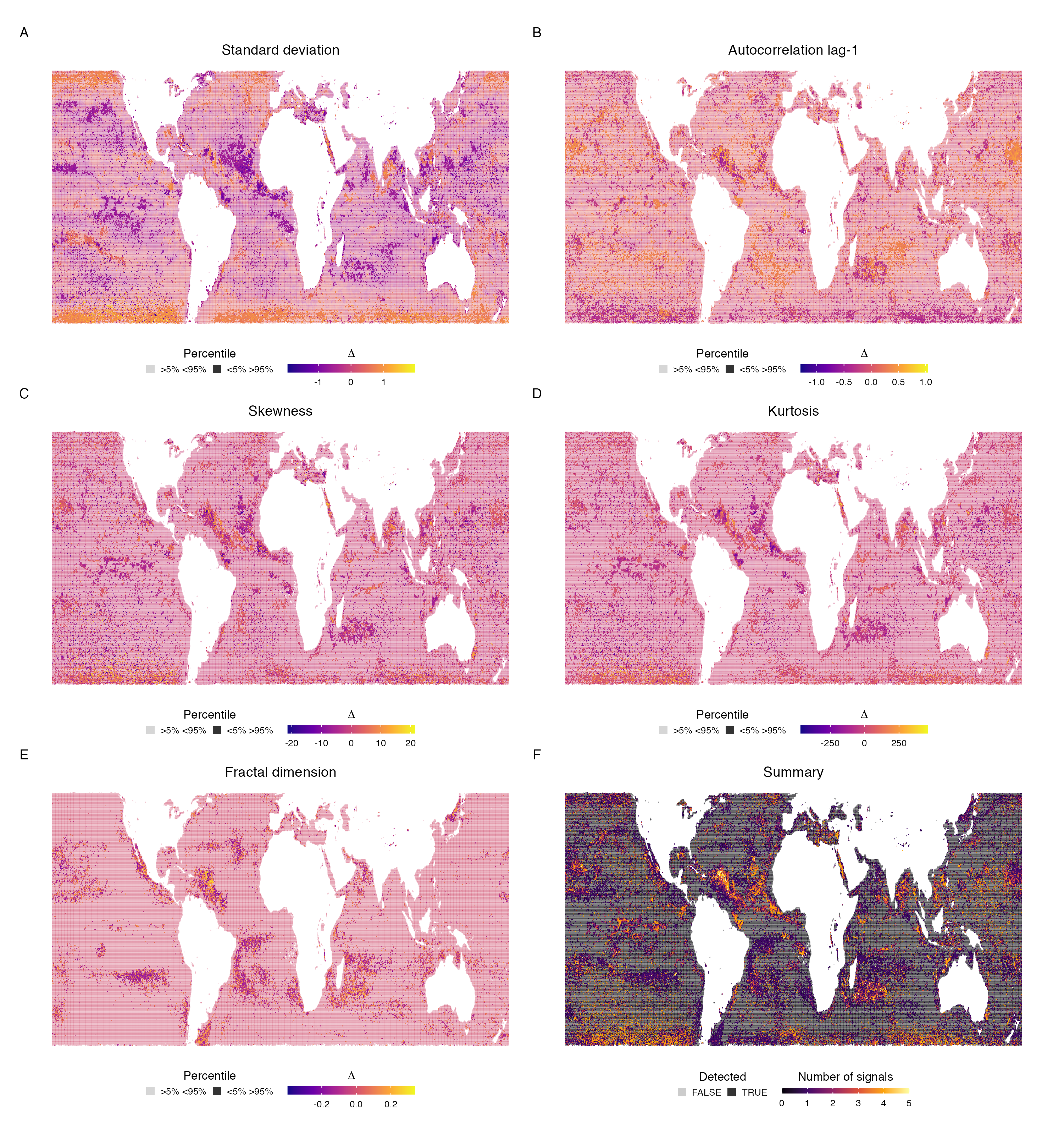}
\caption{\textbf{Resilience indicators in chlorophyll A} Maps of the $\Delta$ outliers per biome are shown for standard deviation (A), autocorrelation at lag-1 (B), skewness (C), kurtosis (D), and fractal dimension (E). Insets show the distribution of $\Delta$ for each biome following the same colouring scheme of fig \ref{fig:mar}. (F) is the summary with the aggregated number of signals.}
\label{fig:sm-mar}
\end{figure*}

\begin{figure*}[ht]
\centering
\includegraphics[width = 6in, height = 5in]{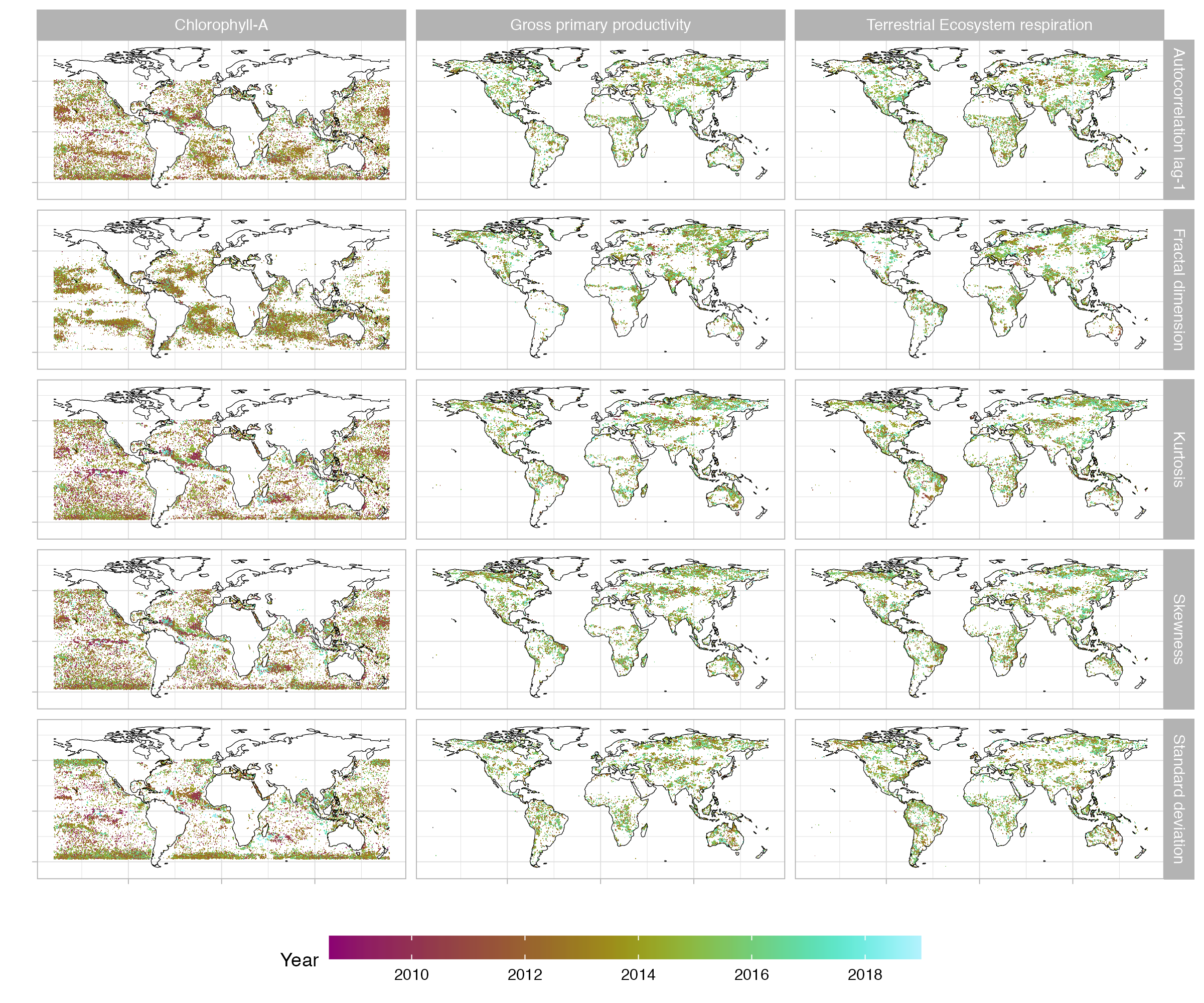}
\caption{\textbf{Spatial and temporal coherence} Break points of the segmented regressions are show as maps for each dataset. The clustering in time and space of the dynamic indicators of resilience supports the idea that some areas are under similar pressures and can shift in tandem.}
\label{fig:sm-map}
\end{figure*}

\begin{figure*}[ht]
\centering
\includegraphics[width = 6in, height = 8in]{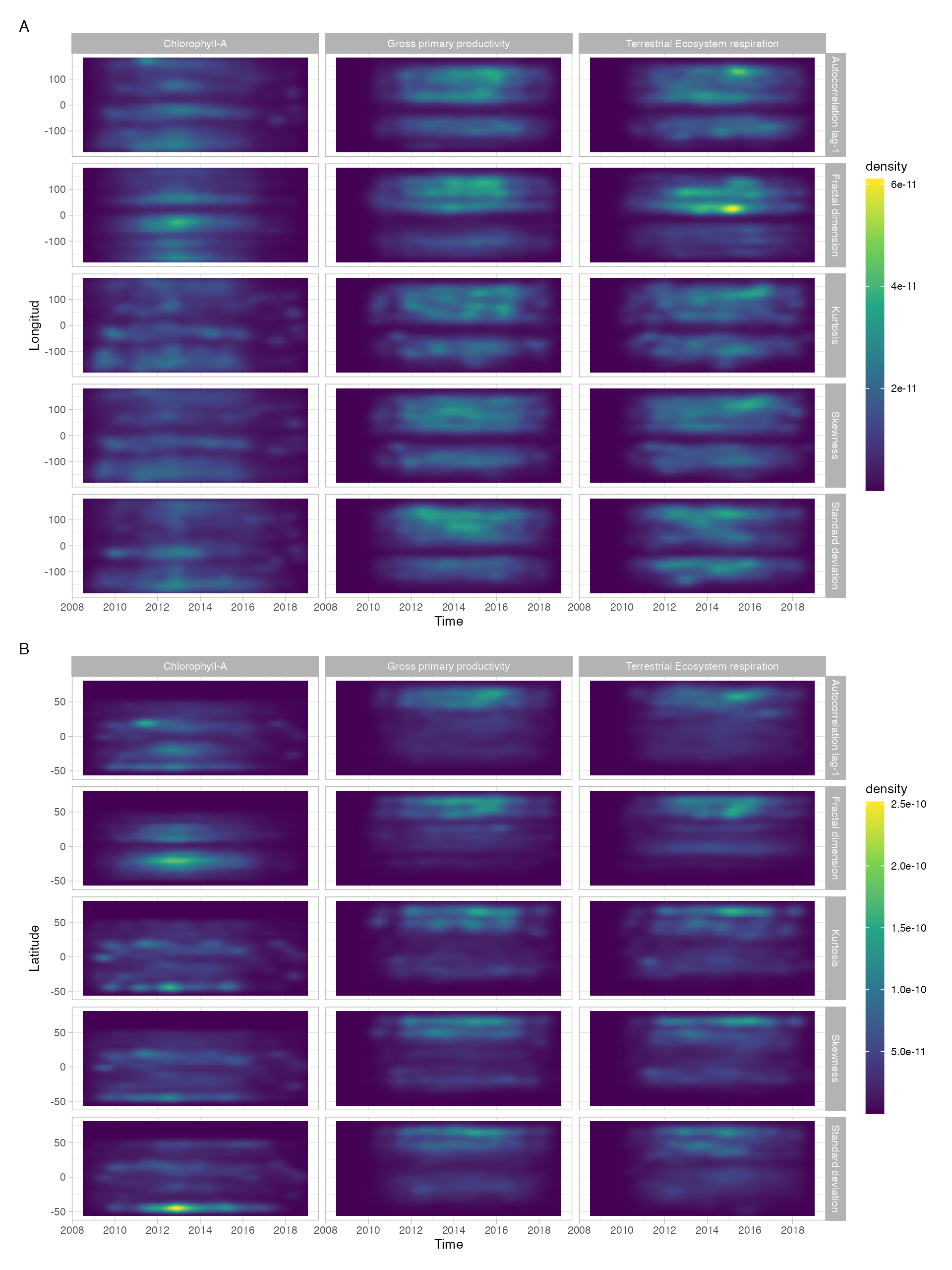}
\caption{\textbf{Temporal coherence of signals} Probability density function of the location of break points in time and space for longitude (A) and latitude (B).}
\label{fig:sm-temp-clus}
\end{figure*}

\begin{figure*}[ht]
\centering
\includegraphics[width = 5in, height = 4in]{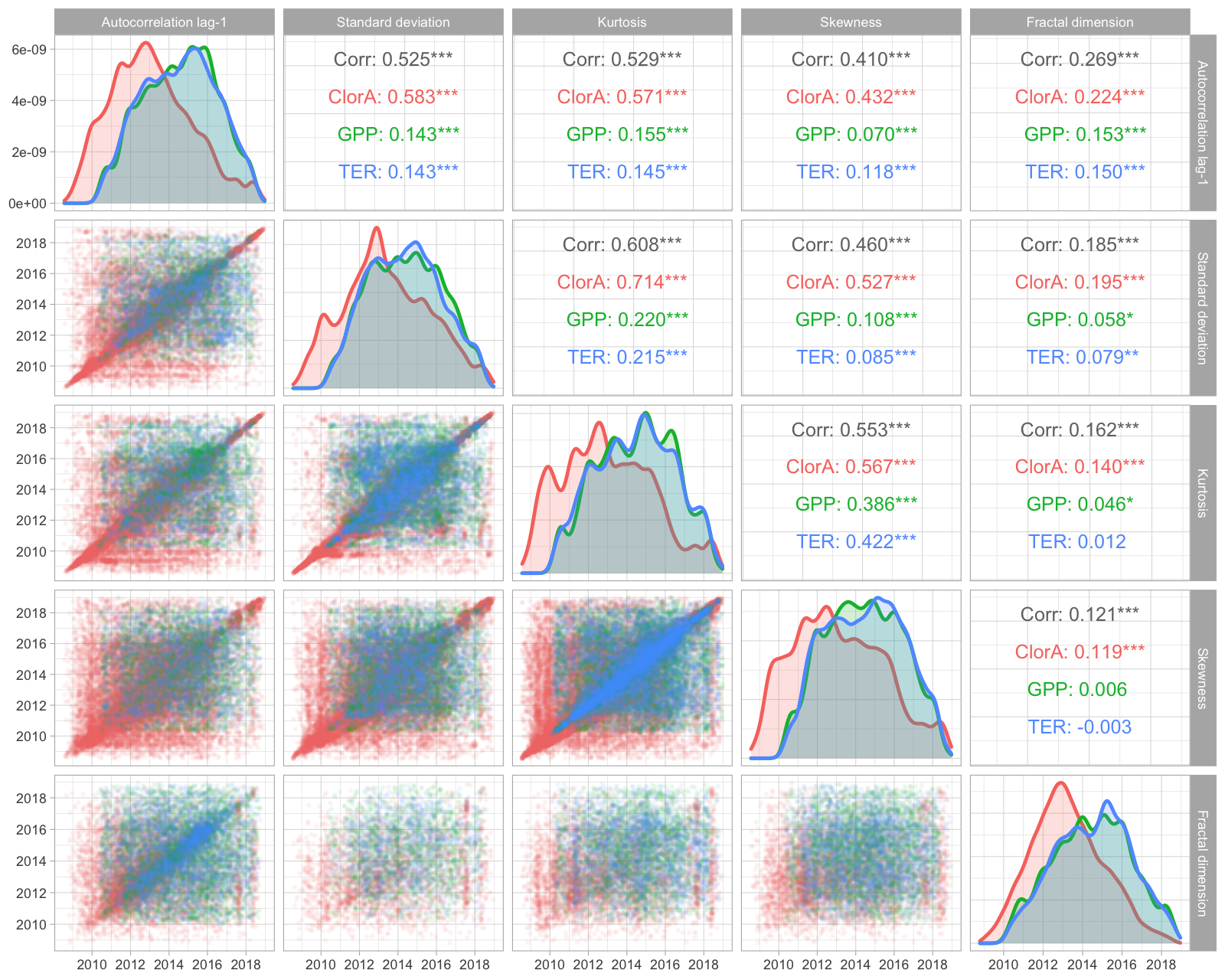}
\caption{\textbf{Temporal correlations} Correlations in time of resilience indicators across datasets: gross primary productivity (GPP), terrestrial ecosystem respiration (TER), and chlorophyll A (ClorA)}
\label{fig:sm-temp-cor}
\end{figure*}

\end{document}